\documentclass[12pt,fleqn]{article}
\usepackage{amsfonts,amssymb,epsfig,latexsym,epic}
\usepackage[vcentermath,enableskew]{youngtab}

\makeatletter
\@addtoreset{equation}{section}
\makeatother

\newcommand{\Ga}{\alpha}
\newcommand{\Gb}{\beta}

\newcommand{\Gd}{\delta}
\newcommand{\Ge}{\epsilon}
\newcommand{\Geps}{\varepsilon}
\newcommand{\Gg}{\gamma}
\newcommand{\GG}{\Gamma}

\newcommand{\Gth}{\theta}
\newcommand{\GTh}{\Theta}

\newcommand{\cro}{\!\times\!}

\newcommand{\pls}{\!+\!}
\newcommand{\mis}{\!-\!}

\newcommand{\cA}{{\scriptscriptstyle\cal A}}
\newcommand{\cB}{{\scriptscriptstyle\cal B}}

\newcommand{\cM}{{\scriptscriptstyle\cal M}}
\newcommand{\cN}{{\scriptscriptstyle\cal N}}
\newcommand{\cK}{{\scriptscriptstyle\cal K}}
\newcommand{\cL}{{\scriptscriptstyle\cal L}}
\newcommand{\cP}{{\scriptscriptstyle\cal P}}
\newcommand{\cQ}{{\scriptscriptstyle\cal Q}}
\newcommand{\cR}{{\scriptscriptstyle\cal R}}

\newcommand{\CD}{{\cal D}}

\newcommand{\CM}{{\cal M}}
\newcommand{\CN}{{\cal N}}

\newcommand{\CL}{{\cal L}}

\newcommand{\CS}{{\cal S}}

\newcommand{\CV}{{\cal V}}

\newcommand{\dA}{{\dot{A}}}

\newcommand{\Be}{{\bar{\Ge}{}}}
\newcommand{\Bchi}{{\bar{\chi}{}}}
\newcommand{\Bpsi}{{\bar{\psi}{}}}
\newcommand{\Bphi}{{\bar{\phi}{}}}

\newcommand{\Bi}{{\bar{\imath}}}

\newcommand{\eqn}[1]{(\ref{#1})}

\newcommand{\bbP}{\mathbb{P}}

\newcommand{\ft}[2]{{\textstyle {\frac{#1}{#2}} }}
\newcommand{\dd}{\partial}

\newcommand{\I}{{\rm i}}

\def\C{\mathbb{C}}

\newcommand{\be}{\begin{equation}}
\newcommand{\ee}{\end{equation}}
\newcommand{\ben}{\begin{displaymath}}
\newcommand{\een}{\end{displaymath}}
\newcommand{\ba}{\begin{eqnarray}}
\newcommand{\ea}{\end{eqnarray}}
\newcommand{\nn}{\nonumber}
\newcommand{\non}{\nonumber\\}
\newcommand{\bea}{\begin{eqnarray}}
\newcommand{\eea}{\end{eqnarray}}
\newcommand{\bean}{\begin{eqnarray*}}
\newcommand{\eean}{\end{eqnarray*}}

\newcommand{\mathon}{\mathversion{bold}}
\newcommand{\mathoff}{\mathversion{normal}}

\def\moth{\mathsurround=0pt}
\newdimen\zo \zo=0pt

\def\tick{\leaders\hrule height 0.5ex depth 0pt \hskip 0.5pt}
\def\upboxfill{$\moth \setbox\zo\hbox{\tick}%
  \hskip 2pt\hbox to 0pt{$\tick$\hss}\hrulefill \hbox to 6pt{$\tick$\hss}$}
\def\underbox#1{\offinterlineskip{\mathord{\mathop{\vtop{\moth\ialign{##\crcr
      $\hfil\displaystyle{#1}\hfil$\crcr\noalign{}
      {\upboxfill}\crcr\noalign{}}}}\limits}}}
\def\dtick{\leaders\hrule height .34pt depth .5ex \hskip 0.5pt}
\def\downboxfill{$\moth \setbox\zo\hbox{\dtick}%
  \hskip 2pt\hbox to 0pt{$\dtick$\hss}\hrulefill \hbox to 6pt{$\dtick$\hss}$}
\def\overbox#1{\mathop{\vbox{\moth\ialign{##\crcr\noalign{}
\downboxfill\crcr\noalign{\vskip 1pt\nointerlineskip}
      $\hfil\displaystyle{#1}\hfil$\crcr}}}\limits}

\def\undersym#1{\underbox{{}#1}}
\def\oversym#1{\overbox{{}#1}}

\makeatletter
\newbox\slashbox \setbox\slashbox=\hbox{$/$}
\def\pFMslash#1{\setbox\@tempboxa=\hbox{$#1$}
  \@tempdima=0.5\wd\slashbox \advance\@tempdima 0.5\wd\@tempboxa
  \copy\slashbox \kern-\@tempdima \box\@tempboxa}
\newcommand\FMslash{\protect\pFMslash}
\makeatother

\newcommand{\SU}[1]{{{\rm SU}({#1})}}
\newcommand{\SO}[1]{{{\rm SO}({#1})}}
\newcommand{\Sp}[1]{{{\rm Sp}({#1})}}

\newcommand{\la}{\label}

\newcommand{\Ref}[1]{(\ref{#1})}

\newcommand{\cc}[1]{g_{\scriptscriptstyle #1}}

\newcommand{\Mat}{L}

\begin{document}

\begin{center}
AEI-2003-105\qquad\quad 
DESY 04-033\qquad\quad 
ITP-UU-04/05\qquad\quad 
SPIN-04/03
\end{center}

\bigskip

\begin{center}

{\bf\Large Gauged Supergravities in Three Dimensions:}
\medskip

{\bf\Large A Panoramic Overview}\footnote{Based on talks by B.~de Wit and
H.~Nicolai at the 27th Johns Hopkins Workshop, 24 - 26 August 2003, 
G\"oteborg, Sweden.}

\end{center}
\setcounter{footnote}{0}

\vspace{4mm}

\begin{center}
{{\bf B.~de Wit}}

{{\em Institute for Theoretical Physics \,\&\, 
Spinoza Institute,\\ 
Utrecht University, Postbus 80.195, 3508 TD Utrecht, 
The Netherlands}}\\
{\small b.dewit@phys.uu.nl} 
\vspace{.5cm}

{{\bf H. Nicolai}}

{{\em Max-Planck-Institut f{\"u}r Gravitationsphysik,\\
  Albert-Einstein-Institut,\\
  M\"uhlenberg 1, D-14476 Potsdam, Germany}}\\
{\small nicolai@aei.mpg.de}

\vspace{.5cm}
{\bf H.~Samtleben}

{{\em II. Institut f\"ur Theoretische Physik der Universit\"at Hamburg,\\ 
 Luruper Chausse 149, D-22761 Hamburg, Germany}}\\
{\small henning.samtleben@desy.de} 

\end{center}

\medskip

\begin{abstract}
Maximal and non-maximal supergravities in three spacetime dimensions allow for
a large variety of semisimple and non-semisimple gauge groups, as well as
complex gauge groups that have no analog in higher   
dimensions. In this contribution we review the recent progress 
in constructing these theories and discuss some of their possible 
applications.
\end{abstract}

\renewcommand{\thefootnote}{\arabic{footnote}}
\setcounter{footnote}{0}
\renewcommand{\thefootnote}{\arabic{footnote}}

\newpage

\section{Introduction}
Locally supersymmetric theories in three spacetime dimensions coupled
to matter have at most $N=16$ supersymmetries \cite{dWToNi93}. The
bosonic matter is described by scalar fields, which parametrize a
target space belonging to a nonlinear sigma model. While there is a large
number 
of possible target spaces when $N\leq 4$, the possibilities become more 
restricted with increasing~$N$: beyond $N=4$, the target spaces are
coset spaces ${\rm G}/{\rm H}$, where ${\rm H}$ is the maximal 
compact subgroup of ${\rm G}$. For all values of $N$ these 
supergravities may be invariant under a bosonic symmetry group ${\rm G}$, 
which commutes with the Lorentz transformations and spacetime 
diffeomorphisms and which involves (subgroups of) the target-space 
isometry group and the R-symmetry group $\SO{N}$. 
In that case there exist supersymmetric deformations where a
subgroup ${\rm G}_0\subset {\rm G}$ is promoted to a local symmetry
\cite{NicSam00,NicSam01b,FiNiSa03,dWHeSa03}, thereby furnishing
three-dimensional analogs of the gauged supergravities in dimensions
$D\geq 4$ that have been known for a long time. In contrast to
higher-dimensional gauged supergravities, the vector fields in general 
appear via a Chern-Simons (CS) rather than a Yang-Mills (YM) term. 
As it turns out, there is a surprisingly rich structure and variety of 
possible gaugings, including semisimple and non-semisimple gauge groups
as well as novel complex gaugings which have no analog in $D\geq 4$
dimensions.
 
There are several reasons why $D=3$ (gauged) supergravities 
are of more general interest. Below we list some of these reasons.

\begin{itemize}
\item During the last five years there has been enormous interest in
the so-called AdS/CFT correspondence, according to which a supergravity 
theory with an AdS groundstate is related to a (super)conformal 
theory living on the boundary of the AdS space (see \cite{AGMOO99} for
a review and an extensive list of references). Much of this interest 
has been focused on the ${\rm AdS}_5/{\rm CFT}_4$ correspondence, 
relating gauged maximal supergravity with gauge group 
${\rm SO(6)}$ on  ${\rm AdS}_5$ \cite{GuRoWa86} to the maximally 
supersymmetric Yang-Mills theory on its boundary. 
While in this case one has essentially only one theory to test the
conjectured correspondence, the number of possibilities is far greater
when one descends by two in the number of dimensions:
the ${\rm AdS}_3 /{\rm CFT}_2$ correspondence offers a much
larger bestiary of examples, because on the one hand there are far more
superconformal theories in two dimensions, and on the other hand because
gauged supergravities are more numerous in three dimensions.

\item (Ungauged) extended supergravities exhibit their maximal global 
and most ``unified'' symmetry in three dimensions\footnote{Here we will 
not be concerned with the {\it infinite dimensional} global symmetries $E_9$,
$E_{10}$ and $E_{11}$, which are known, resp. conjectured, to emerge for 
maximal supergravities in dimensions $D\leq 2$.}, because all tensor gauge 
fields can be dualized to scalar fields, so that the propagating
bosonic degrees of freedom are uniformly described by scalar fields,
which usually live on a target space with a nice geometrical structure.
In particular for the maximal $N=16$ theory, the maximally extended 
exceptional Lie algebra ${\rm E}_8$ makes its appearance 
\cite{Juli83,MarSch83}, whereas in dimensions $D=4$ and $D=5$ the 
maximal-rank exceptional symmetries compatible with maximal supersymmetry 
are ${\rm E}_7$ and ${\rm E}_6$, respectively~\cite{CreJul79}.

\item Unlike the abelian duality relating scalar fields and antisymmetric
tensor gauge fields in higher dimensions, the duality between scalar and 
vector fields can be extended to non-abelian gauge groups in three 
dimensions. There is a novel equivalence between YM and certain CS gauge 
theories (which also holds for non-supersymmetric theories) which
has no analog in dimensions $D\geq 4$. Namely, as shown 
in~\cite{NicSam03a,dWHeSa03}, in three dimensions, any YM gauged 
supergravity with gauge group ${\rm G}_{\rm YM}$ is equivalent to a 
CS gauged supergravity with non-semisimple gauge group 
${\rm G}_{\rm YM}\ltimes {\cal T}$ with a certain translation 
group ${\cal T}$. Because in the latter formulation all the vectors 
appear via a CS term rather than a YM-type kinetic term, no new 
propagating degrees of freedom are generated by the gauging, as is 
required by the preservation of supersymmetry. Altogether, the CS gauged 
supergravities thus not only contain the YM-type gauged theories but
encompass a much larger class of theories.

\item Because the vectors appear via a CS term and do not propagate,
their number and hence the dimension of the gauge group are not determined 
{\it a priori}, unlike in dimensions $D\geq 4$. For this reason, the 
possible gaugings are more numerous and exhibit a richer structure than 
the corresponding $D\geq 4$ gauged supergravities. Similar comments apply 
to the scalar potentials of these theories which provide a large variety 
of symmetry breaking patterns with vacua of the anti-de Sitter, Minkowski 
or de Sitter-type \cite{Fisc02,FiNiSa02,Fisc03}. Among the novel features 
without analog in higher dimensions let us mention the existence of 
maximally supersymmetric vacua for non-compact gauge groups (cf. table~2
in section~9) and the occurrence of stable AdS-type vacua with completely 
broken supersymmetry (for $D\geq 4$, all known non-supersymmetric vacua of 
maximally gauged supergravities are unstable 
\cite{Warn83,deWNic84,HulWar85,Hull85}).

\item Except for certain non-semisimple gaugings, none of the $D=3$ gauged
supergravities can be obtained by any known mechanism from higher 
dimensional supergravity. The very existence of these theories may 
thus point to the existence of new ``cusps'' of M theory, and the existence 
of new geometrical structures in eleven dimensions of the type suggested 
in \cite{KoNiSa00,dWiNic00} and references therein. The theories 
which do originate from higher dimensions usually appear with
a YM-type kinetic term, and therefore necessarily require non-semisimple 
gauge groups in the CS-type formulation, as described above. In particular
they include all those theories obtained by reduction of higher-dimensional 
maximal gauged supergravities on a torus, or by Kaluza-Klein compactification 
of higher-dimensional supergravities on some internal manifold, such as 
for instance IIA/IIB supergravity compactified on the seven-sphere, 
or $D=5$ supergravity on the two-sphere.

\item Just like $D=11$ supergravity can be viewed as a strong-coupling
limit of $D=10$ IIA superstring theory~\cite{Witt95a} one may speculate 
that four dimensions might arise out of a strongly coupled
$D=3$ supergravity theory~\cite{Witt95b}. In this context, a special 
role is played by the dilaton field, whose expectation value on the one hand 
`measures' the size of the $S^1$ on which one reduces, and on the other 
hand appears as the string coupling constant. The connection between
the pertinent ${D=3}$ potentials and the potentials of $D\geq 4$ gauged 
supergravity potentials has been studied in \cite{FiNiSa03}, where the
dilaton is identified with the scalar field associated with a certain
grading operator which is an element of the relevant (non-semisimple) 
gauge group.

\item Gauged supergravity can provide an effective and economical  
description of an {\em infinite} number of Kaluza-Klein supermultiplets in 
a way that is again without analog in dimensions $D\geq 4$. This has been 
recently demonstrated for the compactification of matter-coupled
half maximal $D\!=\!6$ supergravity on ${\rm AdS}_3 \times S^3$ which 
leads to an effective theory in three dimensions with $N=8$ local 
supersymmetries \cite{NicSam03b}. More specifically, the self-interactions 
of the the spin-1 Kaluza-Klein  
towers are fully described by an $N=8$ gauged supergravity with gauge group 
$\SO4\ltimes {\cal T}_\infty$, where ${\cal T}_\infty$ is an infinite 
dimensional translation group, and the gauge group is embedded into the 
global symmetry group $\SO{8,\infty}$. (The second entry is infinite because 
there are infinitely many $N=8$ matter supermultiplets.) In particular,
this embedding is compatible with the quantum numbers of the 
Kaluza-Klein supermultiplets, and the masses of all Kaluza-Klein
states are correctly recovered from a single scalar potential.

\item Finally, there are intriguing connections to recent developments in
the differential geometry of three-dimensional manifolds. On the one 
hand, the models contain the CS Lagrangians that can be used to 
describe knots and links and their characteristic polynomials
(invariants) \cite{Witt91,KasRes02,Guko03}. On the other hand they contain 
the requisite matter fields to realize the various elementary Thurston 
geometries \cite{Gege02,GegKun03}; in particular, recent progress in 
establishing part of the Thurston conjecture \cite{Pere02} has been
based on the introduction of a `dilaton field'. The question is 
therefore whether these gauged supergravities can provide a unified
framework for these so far disconnected parts of mathematics.
\end{itemize}

This review is organized as follows. In section~2 we briefly review the
results of \cite{dWToNi93} on the ungauged supergravity theories in
three dimensions. The global invariances of the corresponding
Lagrangians are discussed in section~3. In section~4 we show that
arbitrary gauge field couplings of Yang-Mills-type in three dimensions 
may always be brought into the form of particular Chern-Simons 
interactions. For general gaugings we may thus restrict attention
to the latter type of theories. In section~5 and~6 we present the full
Lagrangian and transformation rules of the gauged supergravities in
three dimensions, as well as the conditions that must be satisfied in order
that the gauging preserves supersymmetry. The theories for $N\le4$
supersymmetries are discussed in more detail in section~7, while 
sections~8 and~9 focus on the structure of the $N>4$ theories, and 
in particular on the admissible gauge groups for the maximal 
($N=16$) theory. There, we also mention some possible implications
of our results for the AdS$_3$/CFT$_2$ correspondence.

\section{Supergravity coupled to nonlinear sigma models}

In this section we briefly summarize the results of \cite{dWToNi93} 
(for a discussion of the peculiarities of pure gravity in three
space-time dimensions, we refer to \cite{DeJaHo84,DesJac84}).
The fields of the nonlinear sigma model are the target-space   
coordinates~$\phi^i$ and their superpartners~$\chi^i$,  
with $i=1, \ldots, d$; the supergravity fields are
the dreibein $e_\mu{}^a$, the spin-connection field~$\omega_\mu{}^{ab}$
and $N$ gravitino fields~$\psi^I_\mu$ with $I=1,\ldots,N$. The
gravitinos transform under the R-symmetry group ${\rm SO}(N)$, which is
not necessarily a symmetry group of the Lagrangian. 

Since the fields are all massless at this stage, one may assume that
no matter fields other than scalars and spinors are required, because 
helicity is trivial in three dimensions. The scalar fields  parametrize 
a target space endowed with a Riemannian metric $g_{ij}(\phi)$. 
Pure supergravity is topological in three dimensions and exists for an
arbitrary number $N$ of supercharges and corresponding gravitinos
\cite{AchTow86}. Its coupling to a nonlinear sigma model requires the
existence of $N-1$ hermitean, almost complex, structures
$f^{Pi}{}_j(\phi)$, labeled by $P= 2, \ldots,N$, which generate a
Clifford algebra,  
\ba
f^{Pi}{}_k \,f^{Qk}{}_j +  f^{Qi}{}_k\, f^{Pk}{}_j = -2\,
\delta^{PQ}\,\delta^i_j \,.
\ea
{}From the $f^P$ one constructs
$\ft12N(N\mis1)$ tensors $f^{IJ}_{ij}=-f^{JI}_{ij}= -f^{IJ}_{ji}$ that
act as generators for the group ${\rm SO}(N)$,
\ba
f^{PQ}= f^{[P}\,f^{Q]} \,,\qquad f^{1P}= -f^{P1} = f^P \,,
\ea
where, here and  henceforth, $I,J=1,\ldots,N$. 
The $f^{IJ}$ are covariantly constant, both with respect to
the Christoffel and ${\rm SO}(N)$ target-space connections,
$\Gamma_{ij}{}^k$ and $Q_i^{IJ}$, respectively,
\ba
\label{Q-curv}
D^{~}_i(\Gamma,Q)\,f^{IJ}_{jk}\equiv \partial_i f^{IJ}_{jk}
- 2\, \Gamma^{~}_{i[k}{}^l\, f^{IJ}_{j]l}  +2\, 
Q^{K[I}_i\, f^{J]K}_{jk} =0\,.
\ea
The ${\rm SO}(N)$ connections $Q_i^{IJ}(\phi)$
are nontrivial in view of 
\ba
R^{IJ}_{ij}(Q)\equiv \partial^{~}_i Q_j^{IJ} - \partial_j Q_i^{IJ} + 2
Q_i^{K[I}Q_j^{J]K} = \ft12 f^{IJ}_{ij} \;.
\label{fQ}
\ea

For $N=2$ the target space is K\"ahler and $f^{12}$ is
a complex structure. The ${\rm SO}(2)$ holonomy is undetermined. 
For $N=3$, there are three (almost) complex structures $f^{12},
f^{23}$ and $f^{31}$, and the target space is a quaternion-K\"ahler space. 
The case $N=4$ is special: there exists a tensor $J^i{}_j$, defined by  
\ba
J=\ft1{24} \varepsilon^{IJKL} f^{IJ} f^{KL} \;, \qquad J^2 
= {\bf 1}\;,
\ea
which has eigenvalues $\pm1$, commutes with the almost complex 
structures and is covariantly constant. This implies that the target  
space is locally the product of two separate Riemannian spaces of 
dimension $d_\pm$, where $d_++d_-=d$ and $d_\pm$ are both multiples of 
4. These two subspaces are quaternion-K\"ahler and correspond to inequivalent
$N=4$  supermultiplets. For $N=4$ we note the following identity, 
\be
f^{IJ\,ij}\,f^{KL}{}_{\!ij}  = 4 \left(d_+\,\bbP_+^{IJ,KL} +
d_-\,\bbP_-^{IJ,KL}\right) \;,
\ee
with (anti)self-dual projectors, 
\ba
\bbP_\pm^{IJ,KL} &=& \ft12 \delta^{I[K}\,\delta^{L]J} \mp\ft14 
\varepsilon^{IJKL} \;.
\label{P4}
\ea

For $N>2$ the target space is an Einstein space with nontrivial ${\rm 
SO}(N)$ holonomy. The holonomy group is contained in ${\rm 
SO}(N)\times {\rm H}^\prime\subset {\rm SO}(d)$ which must act 
irreducibly on the target space. The group ${\rm H}^\prime$ must be a subgroup
of ${\rm SO}(k)$ (for $N=7,8,9$~mod~$8$), ${\rm U}(k)$ (for
$N=2,6$~mod~$8$), or ${\rm Sp}(k)$ (for $N=3,4,5$~mod~$8$), where $k$ denotes
the number of independent supermultiplets whose scalar fields parametrize the
target space. For $N=4$ these results are more subtle 
because of the product structure. We note the following relations 
(always assuming $N>2$),  
\ba
R_{ijkl} &=& \ft18\,\Big( f^{IJ}_{ij} \,f^{IJ}_{kl} + C_{\alpha\beta}
\,h^\alpha_{ij}\,h^\beta_{kl}\Big) \;, \nn\\ 
R_{ijkl}\,f^{IJ\,kl} &=& \ft12\left(d_+\,\bbP_+{}^{IJ,KL} +
d_-\,\bbP_-{}^{IJ,KL}\right) f_{ij}^{KL}\;, \nonumber\\
R_{ij} &=& (N-2 +\ft18 d)\, g_{ij} + \ft18 (d_+-d_-) J_{ij} \;, 
\label{geom}
\ea
where, for $N\not=4$, one must set $J_{ij}=0$
and $\bbP_\pm{}^{IJ,KL}= \ft12 \delta^{I[K}\delta^{L]J}$. In the first
equation, $C_{\alpha\beta}(\phi)$ is a symmetric tensor and the
target-space tensors $h^\alpha_{ij}(\phi)$ form a basis of
antisymmetric tensors commuting with the almost complex
structures. These tensors generate the ${\rm H}^\prime$ factor of the 
holonomy group with corresponding structure constants $f^{\Ga\Gb}{}_\Gg$.
\begin{table}[bt]
\begin{center} 
\begin{tabular}{||c|c|c||} \hline 
$N$ & Target Space & $d$ 
\\[.5ex] \hline\hline
&&\\[-2.1ex]
1& Riemann manifold $\cal{M}_{\rm R}$
& $k$ 
\\[.5ex] \hline
&&\\[-2.1ex]
2& K\"ahler manifold $\cal{M}_{\rm K}$
& $2k$ 
\\[.5ex] \hline
&&\\[-2.1ex]
3& quaternion K\"ahler manifold $\cal{M}_{\rm QK}$
& $4k$ 
\\[.5ex] \hline
&&\\[-2.1ex]
4&  quaternion K\"ahler manifolds 
$\cal{M}_{\rm QK1}\cro\cal{M}_{\rm QK2}$
& $4(k_1\!+\!k_2)$ 
\\[.5ex] \hline
&&\\[-2.1ex]
5& $\Sp{2,k}/(\Sp{2}\cro\Sp{k})$ 
& $8k$ 
\\[.5ex] \hline
&&\\[-2.1ex]
6& $\SU{4,k}/({\rm SU}(k)\cro{\rm SU}(4)\cro {\rm U}(1))$
& $8k$
\\[.5ex] \hline
&&\\[-2.1ex]
8& $\SO{8,k}/(\SO8\cro\SO{k})$
& $8k$
\\[.5ex] \hline
&&\\[-2.1ex]
9& ${\rm F}_{4(-20)}/\SO9 $
& 16 
\\[.5ex] \hline
&&\\[-2.1ex]
10& ${\rm E}_{6(-14)}/(\SO{10}\cro{\rm U}(1)) $
& 32
\\[.5ex] \hline
&&\\[-2.1ex]
12& ${\rm E}_{7(-5)}/(\SO{12}\cro\Sp1) $
& 64 
\\[.5ex] \hline
&&\\[-2.1ex]
16& ${\rm E}_{8(8)}/\SO{16} $
& 128 
\\[.5ex] \hline
\end{tabular} 
\end{center}
\caption{\small Target spaces for $D=3$ supergravities. The number of
  independent supermultiplets is denoted by $k$. For $N=4$ there exist two
  types of (inequivalent) supermultiplets, counted by $k_1$ and $k_2$. } 
\label{GG1} 
\end{table} 

Beyond $N=4$ the target space geometries become very restricted. This is 
shown in table~\ref{GG1}, where $k$ denotes the number of matter
supermultiplets coupled to supergravity. Remarkably, not all values of
$4<N\leq 16$ can be realized: matter-coupled supergravities exist only for
$N=5,6,8,9,10,12$  and 16 supercharges. Furthermore, only for $N\leq 8$ is it
possible to  include an arbitrary number $k$ of supermultiplets, whereas for
$N\geq 9$ there exists only one theory for each admissible value of $N$.

Let us now turn to the Lagrangian and supersymmetry transformations.
We adopt a manifestly ${\rm SO}(N)$ covariant
notation which allows to select the $N\mis1$ almost complex structures 
from the $f^{IJ}$ tensors by specifying some arbitrary unit $N$-vector 
$\alpha_I$ and identifying the complex structures with $\alpha_J 
f^{JI}$. Accordingly we extend the fermion fields $\chi^i$ to an
overcomplete set, $\chi^{iI}$, defined by
\ba
\chi^{iI} = \left(\chi^i, f^{Pi}{}_j\,\chi^j \right) \;.
\ea

The fact that we have only $d$ fermion fields, rather than $dN$, is 
expressed by the ${\rm SO}(N)$ covariant constraint,  
\ba 
\chi^{iI} &=& \bbP_J^I{}_j^i\,  \chi^{jJ} ~\equiv~
{1\over N}\left(\Gd^{IJ}\Gd^{i}_j - f^{IJ\,i}{}_j\right) \chi^{jJ}
\;.
\label{projchi}
\ea
We should stress, that the introduction of $\chi^{iI}$ is a purely
notational convenience; at every step in the computation one  
may change back to the original notation by choosing $\chi^i =
\alpha_I\chi^{iI}$. The covariant notation does not imply 
that the theory is ${\rm SO}(N)$ invariant; rather the covariant setting
allows us to treat the $N$ supersymmetries and the corresponding 
gravitinos on equal footing.

The Lagrangian then takes the form \ba
\CL_0 &=& -\ft12\,\I\,\varepsilon^{\mu\nu\rho}\left(
e_\mu{}^a\,R_{\nu\rho a} +
\Bpsi{}^I_\mu D_\nu\psi^I_\rho  \right)
-\ft12 e\,g_{ij} \left( g^{\mu\nu}\, \partial_\mu \phi^i\,
\partial_\nu \phi^j + N^{-1} \Bchi{}^{iI} \FMslash{D}
\chi^{jI} \right)\non[1ex]
&&{}+ \ft14 e\, g_{ij}\,  \Bchi{}^{iI}\Gg^\mu\Gg^\nu
\psi_\mu^I\,(\dd_\nu\phi^j+\widehat \partial_\nu \phi^j)  
-\ft1{24} e \,N^{-2} R_{ijkl} \; \Bchi{}^{iI} \gamma_a
\chi^{jI} \, \Bchi{}^{kJ} \gamma^a \chi^{lJ} \non[1ex] 
&&{} +  \ft1{48} e\,N^{-2} \left( 3\,( g_{ij}\, \Bchi{}^{iI}
\chi^{jI})^2 - 2(N-2)\, ( g_{ij}\, \Bchi{}^{iI}\gamma^a 
\chi^{jJ})^2  \right)
\;,
\label{L}
\end{eqnarray}
with the covariant derivatives
\begin{eqnarray}
  D_\mu \psi^I_\nu &=& \left( \partial_\mu +\ft12
\omega^a_\mu \,\gamma_a \right ) \psi^I_\nu + 
\partial_\mu \phi^i \, Q_i^{IJ} \psi^J_\nu\;, \non[.5ex]
  D_\mu \chi^{iI} &=&  
\left( \partial_\mu +\ft12 \omega^a_\mu\, \gamma_a \right)\chi^{iI} 
+\partial_\mu \phi^j \left( \Gamma^i_{jk}\, \chi^{kI} +
Q_j^{IJ} \chi^{iJ} \right)  \;. 
\end{eqnarray}
As in \cite{dWToNi93,dWHeSa03},  we use the Pauli-K\"all\'en metric with 
hermitean gamma matrices $\gamma^a$, satisfying  
$\Gg_a\Gg_b = \delta_{ab} +\I \Geps_{abc} \Gg^c$.  
The Lagrangian is invariant under the following supersymmetry transformations
\begin{eqnarray}
\Gd e_\mu{}^a &=& \ft12\,\Be{}^I\Gg^a\,\psi^I_\mu 
\;,\non
\Gd \psi^I_\mu &=& D_\mu \epsilon^I   - \ft18 g_{ij} \,\Bchi{}^{iI}
\gamma^\nu \chi^{jJ}\, \gamma_{\mu\nu} \,\epsilon^J - 
\Gd \phi^i \,Q_i^{IJ} \psi^J_\mu \;,\non
\Gd \phi^i &=& \ft12\,\Be^I\,\chi^{iI} 
\;,\non
\Gd \chi^{iI} &=& \ft12 \left(\delta^{IJ}{\bf 1} \mis f^{IJ}
\right)^i{}_{\!j}\; 
\FMslash{\widehat\partial} \phi^j \, \epsilon^J -
\Gd \phi^j \left(
\Gamma^i_{jk}\,\chi^{kI} 
+ Q_j^{IJ}\chi^{iJ} \right)
\;,
\label{susytraf}
\end{eqnarray}
with the supercovariant derivative
 $\widehat\partial_\mu \phi^i\equiv\partial_\mu \phi^i -\ft12
\Bpsi{}^I_\mu \chi^{iI}$.
Observe that the terms proportional to $\delta \phi$  in 
$\delta\chi^{iI}$ do not satisfy the same constraint
\eqn{projchi} as $\chi^{iI}$ itself, because the projection operator 
$\bbP_J^I{}_j^i$ itself transforms under supersymmetry, such that only
the projector condition is supersymmetric.

\section{Isometries and R-symmetries}

The Lagrangian \Ref{L} and the transformation
rules \Ref{susytraf} are consistent with target-space diffeomorphisms
and field-dependent  ${\rm SO}(N)$ R-symmetry rotations. These transformations
correspond to repara\-metrizations
within certain equivalence classes, but do not, in general, constitute
an invariance. The \SO{$N$} rotations act on $\psi^I_\mu$,
$\chi^{iI}$  and $Q_i^{IJ}$ according to 
\ba
\delta\psi^I_\mu = {\cal S}^{IJ}(\phi)\, \psi^J_\mu\,,\qquad 
\delta\chi^{iI} = {\cal S}^{IJ}(\phi)\, \chi^{iJ}\,,\qquad 
\delta Q_i^{IJ} = - D_i {\cal S}^{IJ}(\phi)\,.
\label{so-1}
\ea
{}From \eqn{fQ}, one concludes that the $f^{IJ}$ should be rotated
correspondingly,
\ba
\delta f^{IJ} = 2\,{\cal S}^{K[I}(\phi) \,f^{J]K}\,.
\label{so-2}
\ea

The bosonic invariance group ${\rm G}$ of the Lagrangian \Ref{L} that commutes
with the Lorentz transformations and spacetime diffeomorphisms, is a subgroup
of the product of the target-space isometries times the R-symmetry
transformations. It is generated by those
target-space isometries whose action on the $Q_i^{IJ}$ and $f^{IJ}$ may
be absorbed by a special ${\rm SO}(N)$ transformation \eqn{so-1}, 
\eqn{so-2}. Specifically, its generators are Killing vector fields $X^i(\phi)$
satisfying  
\ba
\CL_{X}\, g_{ij} ~=~  0 \;, \qquad
\CL_{X} Q_i^{IJ} + D_i {\cal S}^{IJ}(\phi,X)&=& 0 
\;,
\non[.5ex]
\CL_{X} f_{ij}^{IJ} - 2\, {\cal S}^{K[I}(\phi,X)\,f^{J]K}_{ij} 
&=& 0 
\;,\label{isomf}
\ea
where  ${\cal S}^{IJ}(\phi,X)$ is the parameter of an infinitesimal
${\rm SO}(N)$ rotation which depends both on $X^{i}(\phi)$ and on the
scalar fields.  
The Lagrangian \Ref{L} is then invariant under the combined
transformations, 
\ba
\label{iso-trans1}
\Gd\phi^i      = X^i(\phi) \;,\quad
\Gd \psi^I_\mu = {\cal S}^{IJ}(\phi,X) \,\psi_\mu^J \;,\quad
\Gd \chi^{iI}  = \chi^{jI}\partial_jX^i +
{\cal S}^{IJ}(\phi,X) \, \chi^{iJ} \;.
\ea
The fermion transformations can be rewritten covariantly,  
\ba
\label{iso-trans2}
\Gd \psi^I_\mu &=& {\cal V}^{IJ}(\phi,X) \,\psi_\mu^J - \Gd\phi^i\,
Q_i^{IJ} \psi^J_\mu \;,\non
\Gd \chi^{iI}  &=& D_jX^i \,\chi^{jI}+
{\cal V}^{IJ}(\phi,X) \, \chi^{iJ} -
\Gd \phi^j \left(\Gamma^i_{jk}\,\chi^{kI} 
+ Q_j^{IJ}\chi^{iJ} \right) \;,
\ea
where $\CV^{IJ}(\phi,X)\equiv X^j Q_j^{IJ}(\phi) +
{\cal S}^{IJ}(\phi,X)$. 
Using \eqn{fQ} and \eqn{Q-curv}, one verifies 
that the second equation of \Ref{isomf} corresponds to,  
\ba
D_i \CV^{IJ}(\phi,X) &=& \ft12 f^{IJ}_{ij}(\phi)\, X^j(\phi) \;,
\label{DXS}
\ea
which shows that $\CV^{IJ}(\phi,X)$ can be 
regarded as as the moment map associated with the isometry $X^i$.  
After contracting \eqn{DXS}  with $f^{MN\,ij}$, one obtains 
\ba 
f^{IJ\,ij}\,D_iX_j  &=& \left\{
\begin{array}{ll}
\ft12 d\,\CV^{IJ} \;, & \mbox{for $N\not= 2,4$} \\[1ex]
(d_+\,\bbP_+^{IJ,KL} + d_-\,\bbP_-^{IJ,KL})
\CV^{KL}\;,  & \mbox{for $N=4$}
\end{array}
\right.
\label{S}
\ea
The last equation of \Ref{isomf} coincides with the integrability condition
related to \Ref{DXS} and is thus automatically satisfied. 

For $N>2$, the above analysis shows that there are no obstructions 
for extending an isometry to an invariance of the Lagrangian.
For $N=2$ this is different: $\CV^{IJ}$ is determined by \Ref{DXS} up to an
integration constant related to the invariance of the Lagrangian
under constant $\SO2$ transformations of the fermions. The isometries leave
the complex structure invariant and are therefore holomorphic. 
For $N=4$ the (anti)selfdual almost complex structures
$\bbP_\pm^{IJ,KL}\,f^{KL}$ live in the corresponding $d_\pm$-dimensional 
quaternion-K\"ahler subspace. The same holds for the moment maps,  
$\bbP_\pm^{IJ,KL}\,\CV^{KL}$, which according to \eqn{DXS} depend only on the
corresponding subspace coordinates. Note, however, that when one of the
subspaces is trivial, say when $d_-=0$, then $\bbP_-^{IJ,KL}\,\CV^{KL}$
corresponds to a triplet of arbitrary constants. This is a consequence of the
fact that the model in this case has a rigid $\SO3$ invariance acting 
exclusively on the fermions.

These integration constants in ${\cal V}^{IJ}$ correspond to the so-called
Fayet-Iliopoulos (FI) terms that are known from the gaugings of
four-dimensional $N=1$ and $N=2$ supergravity. Indirectly, the above results
may have implications for higher-dimensional gauged supergravities, as follows
from considering 
their reduction to three dimensions. For instance, the reduction of  $d=4, N=1$
supergravity leads to ${d=3}, {N=2}$ supergravity for which the moment maps
can always be modified by an additive constant. Consequently, we expect that 
there are no obstructions against a FI term in four dimensions, which is 
indeed the case. 
For $d=4,N=2$ supergravity the situation is more subtle. The
reduction of these theories to three dimensions leads to a product of two
quaternion-K\"ahler target spaces, one
associated with the vector multiplets and one associated with the
hypermultiplets in four dimensions. As in three dimensions there are no  
integration constants in the moment maps unless one of these
quaternion-K\"ahler spaces is of dimension zero, it follows that 
FI terms are only possible in four dimensions in the absence of
hypermultiplets, a result which is indeed well known.

The generators of ${\rm G}$ are labeled by indices $\CM,\CN\ldots$ and
generate an algebra ${\mathfrak g}$. They consist of combined isometries
generated by Killing vectors $X^{\cM i}$ and infinitesimal ${\rm SO}(N)$ rotations 
${\cal S}^{\cM\,IJ}\equiv{\cal S}^{IJ}(\phi, X^{\cM})$. For $N=2,4$ one may
have the situation that some of the $X^\cM$ vanish, while the
corresponding ${\cal S}^{\cM\,IJ}$ are constant. Closure of ${\mathfrak g}$
implies, 
\ba 
X^{\cM j}\,\dd_j X^{\cN i} - X^{\cN j}\,\dd_j X^{\cM i}  &=&
f^{\cM\cN}{}_{\!\!\cK}\,X^{\cK i} \;,
\label{fABC}\\
\label{SABC}
  \left[{\cal S}^{\cM}, {\cal S}^{\cN}\right]^{IJ} 
- X^{{\cM}i} \,\partial_i {\cal S}^{\cN\,IJ} +
X^{{\cN}i} \,\partial_i {\cal S}^{\cM\,IJ} &=&
- f^{{\cM}{\cN}}{}_{\!\!\cK} \,{\cal S}^{\cK\,IJ} \;,
\ea
with structure constants $f^{\cM\cN}{}_{\!\!\cK}$. 

{From} the integrability condition of \eqn{DXS} one derives that $D_i 
X_j - \ft14 \,f^{MN}_{ij}\,\CV^{MN}$ commutes with the almost complex
structures. For $N>2$ this implies that it can be decomposed in terms of the
antisymmetric tensors $h^\alpha_{ij}$ introduced in \eqn{geom},
\ba
D_i X^\cM_j - \ft14
f^{IJ}_{ij}\,\CV^{\cM\,IJ} &\equiv& h^\alpha_{ij}\,\CV^\cM{}_\alpha
\;.
\label{V2}
\ea
Introducing the notation $\CV^{\cM
\,i}\equiv X^{\cM\,i}$, we establish the following system of linear
differential equations,
\ba
D_i \CV^{\cM\,IJ} &=& \ft12\,f^{IJ}_{ij}\,\CV^{\cM\, j} \;,
\non[.5ex]
D_i  \CV^\cM{}_{\alpha} &=&
\ft18\,C_{\alpha\beta}\,h^\beta_{ij}\,\CV^{\cM\,j}  \;,
\non[.5ex]
D_i  \CV^\cM{}_{j} &=& \ft14\,f^{IJ}_{ij}\,\CV^{\cM\,IJ} 
+ h^\alpha_{ij}\,\CV^\cM{}_\alpha \;,
\label{DV}
\ea
where the covariant derivative contains the Christoffel connection as
well as the ${\rm SO}(N)\times {\rm H}^\prime$ connections. 
Furthermore, we derive
\ba
\label{ABCH}
f^{{\cM}{\cN}}{}_{\!\!\cK} \,\CV^{\cK\,IJ} &=& 
\ft12\,f^{IJ}_{ij}\,\CV^{\cM i}\CV^{\cN j} -
\left[ \CV^\cM  ,\, \CV^\cN \right]^{IJ} 
\;,\non[.5ex]
f^{\cM\cN}{}_\cK\, \CV^\cK {}_\alpha &=&  
\ft18 C_{\alpha\beta} \, h^\beta_{ij}
\,\CV^{\cM\,i} \,\,\CV^{\cN\,j} +
f^{\beta\gamma}{}_\alpha \,
\CV^\cM{}_\beta\, \CV^\cN{}_\gamma  
   \;,\non[.5ex]
f^{\cM\cN}{}_\cK\, \CV^{\cK}{}_i &=& \ft 14 f^{IJ}_{ij} 
(\CV^{\cM\,IJ} \,\CV^{\cN\,j}- \CV^{\cN\,IJ} \,\CV^{\cM\,j})  
+h^{\alpha}_{ij} 
(\CV^{\cM}{}_\alpha \,\CV^{\cN\,j}- \CV^{\cN}{}_\alpha
\,\CV^{\cM\,j})\;.
\ea
Under the ${\rm G}$-transformations the quantities $\CV^{\cM\,IJ}$,
$\CV^{\cM\,i}$ and $\CV^{\cM}{}_\alpha$ 
transform according to the adjoint representation of ${\rm G}$, up to
field-dependent ${\rm SO}(N)\times {\rm H}^\prime$ 
transformations, as is shown by,
\ba
\label{cov-CV}
\CV^{\cN\,i}D_i\CV^{\cM\,IJ} &=& - f^{\cM\cN}{}_\cK\, \CV^{\cK\,IJ} +
[\CV^\cN,\CV^\cM]^{IJ}\;, \non
\CV^{\cN\,i}D_i \CV^{\cM}{}_\alpha &=& - f^{\cM\cN}{}_\cK\,
\CV^{\cK}{}_\alpha  + f^{\beta\gamma}{}_\alpha \,\CV^\cN{}_\gamma\,
\CV^\cM{}_\beta \;.
\ea
For $\CV^{\cM\,i}$ this result is captured by \eqn{fABC}.

\section{Yang-Mills versus Chern-Simons gauged theories}
So far we have been concerned with massless matter fields. We now turn to  
supersymmetric deformations of these theories that can be obtained by
gauging. In that case two issues arise immediately. First the theories
discussed so far did not include vector fields that are obviously needed to
effect the gauging. Secondly, when the fields are not massless then it is no 
longer obvious that matter supermultiplets can be exclusively 
described in terms of scalar and spinor fields, and one might want to 
include other fields as well. As it turns out, these 
two issues are somewhat related.  

First of all, one can always include vector gauge fields without 
changing the number of dynamic degrees of freedom, by introducing 
CS terms. This seems to leave open the option of adding 
additional standard YM kinetic terms (which may eventually 
acquire mass terms by spontaneous symmetry breaking) to describe some 
of the matter degrees of freedom. In fact, all the theories 
that have been constructed by direct dimensional reduction appear 
as YM rather than CS gauged theories~\cite{CvLuPo00,LuPoSe02}. 

However, it turns out that the YM 
Lagrangians in three dimensions are simply equivalent to particular CS 
Lagrangians. The dynamic degrees of freedom are then carried by extra 
(compensating) scalar fields. In this conversion every 
gauge field is replaced by two gauge fields and a new scalar field,
which together describe the same number of dynamic degrees of freedom
as the original gauge field. The 
nonabelian gauge group is enlarged to a bigger gauge group which is necessarily
non-semisimple. 
To see how this comes about, consider a Lagrangian in three spacetime 
dimensions with YM kinetic terms quadratic in the field strengths and with
moment interactions proportional to gauge covariant tensors 
$O^A_{\mu\nu}$,
\ba
\label{YM-L}
\CL = -\ft14\sqrt{g}
\left(F^A_{\mu\nu}(A)\pls O^A_{\mu\nu}(A,\Phi)\right)M_{AB}(\Phi)
\left(F^{B\mu\nu}(A)\pls O^{B\mu\nu}(A,\Phi)\right)  +
\CL^\prime(A, \Phi) \;.
 \ea
Here $A_\mu^A$ denote the nonabelian gauge fields
labeled by indices $A,B,\ldots$,
and $\Phi$ generically denotes possible matter fields transforming
according to certain representations of the gauge group ${\rm G}_{\rm
YM}$. The structure constants of this group are denoted by
$f_{AB}{}^{\!C}$,  so that the field strengths read, 
\ba
F^A_{\mu\nu}(A) = \partial_\mu A_\nu^{A} - \partial_\nu A_\mu^{A}
-f_{BC}{}^{\!A}\, A_\mu^{B} A_\nu^{C} \;. 
\nn
\ea
The symmetric matrix $M_{AB}(\Phi)$ may depend on the matter fields
and transforms covariantly under ${\rm G}_{\rm YM}$. The last term,
$\CL^\prime(A,\Phi)$, in the Lagrangian is separately gauge invariant
and its dependence on the gauge fields is exclusively contained in
covariant derivatives of the matter fields or in topological mass
terms ({\it i.e.}\ CS terms). 

Usually the duality is effected by regarding the field strength as an
independent field on which the Bianchi identity is imposed by means of
a Lagrange multiplier. Because the Lagrangian \eqn{YM-L} depends
explicitly on both the field strengths and on the gauge fields, we
proceed differently and write the field strength in terms of new
vector fields $B_{A\,\mu}$ and the derivative of compensating scalar
fields $\phi_A$, all transforming in the adjoint representation of the
gauge group. The explicit expression,
\ba
\label{F-B}
\ft12\, \I\, \sqrt{g}\, \varepsilon_{\mu\nu\rho}
( F^{A\,\nu\rho}(A) +O^{A\,\nu\rho}(A,\Phi)) 
&=& M^{AB}(B_{B\,\mu} - D_\mu \phi_B)\;,
\ea
where $M_{AC}\,M^{CB}= \delta^B_A$, should be regarded as a field
equation that follows from the new Lagrangian \Ref{LCS} that we are about to
present. The structure of \eqn{F-B} implies that we are dealing with 
additional gauge transformations as its right-hand side is invariant
under the combined transformations, 
\ba
\delta B_{A\,\mu} = D_\mu \Lambda_A \,,\qquad \delta \phi_A =
\Lambda_A \;,
\ea
under which all other fields remain invariant. The corresponding
abelian gauge group, $\cal T$, has nilpotent generators transforming
in the adjoint representation of ${\rm G}_{\rm YM}$. Obviously, the
$\phi_A$ act as compensating fields with respect to $\cal
T$. The combined gauge group is now a semidirect product of ${\rm
G}_{\rm YM}$ and $\cal T$ and its dimension is twice the dimension of 
the original gauge group 
${\rm G}_{\rm YM}$. The covariant field strengths belonging to the new
gauge group are 
$F^A_{\mu\nu}{}(A)$ and $F_{A\,\mu\nu}(B,A)=
2\,D_{[\mu}B_{A\,\nu]}$, and transform under $\cal T$ 
according to $\delta F^A_{\mu\nu}{}=0$ and $\delta F_{A\,\mu\nu}=
- \Lambda_C \, f_{AB}{}^{\!C}\,F^B_{\mu\nu}{}$. The fully gauge
covariant derivative of $\phi_A$ equals 
\ba
{\hat D}_\mu \phi_A &\equiv&
D_\mu\phi_A  - B_{A\,\mu} ~=~  \partial_\mu\phi_A -
f_{AB}{}^{\!C} \,A^B_\mu{}\,\phi_C  - B_{A\,\mu}\;,
\ea
and is invariant under ${\cal T}$ transformations. 

The field equations corresponding to the new Lagrangian,
\ba
\CL &=& -\ft12 \sqrt{g}\,\hat{D}_\mu \phi_A
\,M^{AB}(\Phi)\hat{D}^\mu \phi_B 
+\ft12 \,\I\, \varepsilon^{\mu\nu\rho} \,
(F^A_{\mu\nu}B_{A\,\rho} -
O^{A}_{\mu\nu}\, \hat{D}_\rho \phi_A )
+ \CL^\prime(A, \Phi) \;, 
\label{LCS}
 \ea
lead to \eqn{F-B} as well as to the same field equations
as before for the matter fields $\Phi$. Observe that the Lagrangian is
fully gauge invariant up to a total derivative. In this way, the YM
Lagrangian has now been converted to a CS Lagrangian, with a
different gauge group and a different scalar field content, although
the theory is still equivalent on-shell to the original one. To obtain
the original Lagrangian \eqn{YM-L}, one simply imposes the gauge
$\phi_A=0$ and integrates out the fields $B_{A\,\mu}$. 

In figure~1 we schematically illustrate the implications of this
equivalence for gauged supergravities in three dimensions. When 
descended from higher dimensions, the ungauged theories usually 
appear with the physical bosonic degrees of freedom in different
guises, as scalar and vector fields. In order to exhibit possible
hidden symmetries, one then dualizes the vector fields after which 
all bosonic degrees of freedom are represented by scalar fields (in
principle there can also exist `intermediate' versions).  The 
resulting theory is then on-shell equivalent to the original one.
Both theories can be gauged, but as the table shows, the on-shell
equivalence persists.

\begin{figure}[bt]
  \begin{center}
    \leavevmode
\begin{picture}(0,0)%
\includegraphics{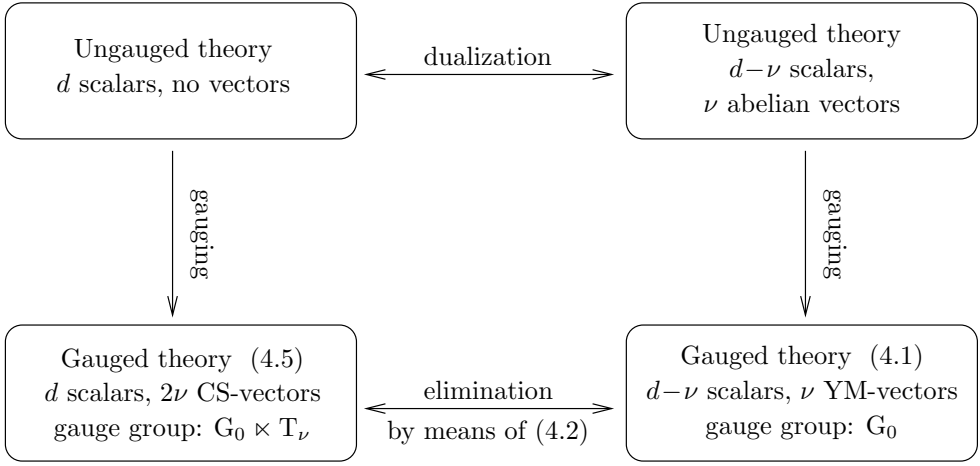}%
\end{picture}%
\setlength{\unitlength}{3947sp}%
\begingroup\makeatletter\ifx\SetFigFont\undefined%
\gdef\SetFigFont#1#2#3#4#5{%
  \reset@font\fontsize{#1}{#2pt}%
  \fontfamily{#3}\fontseries{#4}\fontshape{#5}%
  \selectfont}%
\fi\endgroup%
\begin{picture}(6129,2904)(1789,-4648)
\put(3001,-2911){\rotatebox{270.0}{\makebox(0,0)[lb]{\smash{\SetFigFont{10}{12.0}{\rmdefault}{\mddefault}{\updefault}gauging}}}}
\put(6976,-2911){\rotatebox{270.0}{\makebox(0,0)[lb]{\smash{\SetFigFont{10}{12.0}{\rmdefault}{\mddefault}{\updefault}gauging}}}}
\put(1951,-4246){\makebox(0,0)[lb]{\smash{\SetFigFont{10}{12.0}{\rmdefault}{\mddefault}{\updefault}\begin{tabular}{c}Gauged theory\,\, \Ref{LCS}\\$d$ scalars, $2\nu$ CS-vectors\\gauge group: ${\rm G}_0\ltimes {\rm T}_\nu$\end{tabular}}}}
\put(5931,-2206){\makebox(0,0)[lb]{\smash{\SetFigFont{10}{12.0}{\rmdefault}{\mddefault}{\updefault}\begin{tabular}{c}\phantom{M}Ungauged theory\\\phantom{M}$d\!-\!\nu$ scalars,\\\phantom{M}$\nu$ abelian vectors\end{tabular}}}}
\put(5731,-4231){\makebox(0,0)[lb]{\smash{\SetFigFont{10}{12.0}{\rmdefault}{\mddefault}{\updefault}\begin{tabular}{c}Gauged theory\,\, \Ref{YM-L}\\$d\!-\!\nu$ scalars, $\nu$ YM-vectors\\gauge group: ${\rm G}_0$\end{tabular}}}}
\put(4431,-2131){\makebox(0,0)[lb]{\smash{\SetFigFont{10}{12.0}{\rmdefault}{\mddefault}{\updefault}dualization}}}
\put(2026,-2191){\makebox(0,0)[lb]{\smash{\SetFigFont{10}{12.0}{\rmdefault}{\mddefault}{\updefault}\begin{tabular}{c}Ungauged theory\\$d$ scalars, no vectors\end{tabular}}}}
\put(4431,-4231){\makebox(0,0)[lb]{\smash{\SetFigFont{10}{12.0}{\rmdefault}{\mddefault}{\updefault}elimination}}}
\put(4201,-4486){\makebox(0,0)[lb]{\smash{\SetFigFont{10}{12.0}{\rmdefault}{\mddefault}{\updefault}by means of \Ref{F-B}}}}
\end{picture}
\end{center}
\caption{\small CS and YM gauged supergravity in three dimensions}
\label{diag}
\end{figure}


Perhaps it is worth pointing out that introducing a mass term to a CS theory
or a YM theory has different consequences with regard to the degrees of
freedom \cite{DesYan89}. An abelian CS term with a regular mass term 
proportional to $\ft12 m\, A_\mu^{\,2}$ yields the following massive wave 
equation for the vector field,
\be
 \partial_\mu A_\nu -\partial_\nu A_\mu= \pm\,
 \I\,m\,\varepsilon_{\mu\nu\rho}\,A^\rho  \;,
\ee
which describes massive degrees of freedom with spin only equal to $+1$ or
$-1$, depending on the sign of the mass term. In
contradistinction, a YM kinetic term with a regular mass term $\ft12 m^2\,
A_\mu^{\,2}$ leads to massive degrees of freedom 
carrying {{\em both}} spin $+1$ and spin $-1$. This doubling of degrees of
freedom is consistent with the YM-CS conversion described above, as a YM
theory takes the form of a CS theory with twice the number of vector fields.

\section{The embedding tensor}
We now wish to deform the Lagrangian \Ref{L} such that it becomes
invariant under a subset of transformations \Ref{iso-trans1} with
spacetime dependent parameters. The corresponding 
subalgebra ${\mathfrak g}_0\subset{\mathfrak g}$ is characterized 
by an embedding tensor $\Theta_{\cM\cN}$ via
\be
\label{subgroup}
X^i = g\,\Theta_{\cM \cN}\,\Lambda^\cM(x)\, X^{\cN \,i}\,,\qquad
\CS^{IJ} =  g\,\Theta_{\cM \cN}\,\Lambda^\cM(x)\, \CS^{\cN IJ}\,,
\ee
with gauge parameters $\Lambda^\cN(x)$ depending on the spacetime
coordinates, and a gauge coupling constant $g$. 
Unless the gauge group ${\rm G}_0$ coincides with the full symmetry group
${\rm G}$,  
the embedding tensor acts as a projector which reduces the number 
of independent parameters according to
\be
{\dim} \, \mathfrak{g}_0 = {\rm rank} \,\Theta \;.
\ee
Although it is not obvious from the way $\Theta_{\cM\cN}$ appears in
\Ref{subgroup} we will see below that it must be gauge invariant and 
symmetric under interchange of the indices $\CM$ and $\CN$. Via 
$t^\cM \Theta_{\cM\cN} t^\cN$ it defines an element in the symmetric 
tensor product $(\mathfrak{g}\otimes \mathfrak{g})_{\rm sym}$.
In order that the gauge tranformations generate a group,
$\Theta_{\cM\cN}$ must satisfy the condition,
\be
\label{embedding-cond}
\Theta_{\cM\cP} \,\Theta_{\cN\cQ}\,f^{\cP\cQ}{}_\cR = \hat
f_{\cM\cN}{}^\cP\, \Theta_{\cP\cR}\,,
\ee
for certain constants $\hat f_{\cM\cN}{}^\cP$, which are subsequently
identified as the structure constants of the gauge group. 
One can verify that the validity of the Jacobi
identity for the gauge group structure constants follows directly from
the Jacobi identity associated with the group ${\rm G}$, subject to projection
by the embedding tensor. The symmetry and  gauge invariance of
$\Theta_{\cM\cP}$ implies  
that $\hat f_{\cM\cP}{}^\cQ\, \Theta_{\cQ\cN}+ \hat f_{\cN\cP}{}^\cQ\,
\Theta_{\cM\cQ} = 0$, which can be written in ${\rm G}$-covariant form, 
\ba
  \Theta_{\cP\cL}\left(f^{\cK\cL}{}_\cM \Theta_{\cN\cK}
+ f^{\cK\cL}{}_\cN \Theta_{\cM\cK}\right) &=& 0 \;. 
\label{subgrouptheta}
\ea

Subsequently we introduce the gauge fields $A_\mu^\cM$ into the
definition of the covariant derivatives. For example, we have 
\be
\CD_\mu\phi^i = \partial_\mu \phi^i +g\, \Theta_{\cM\cN} \,A^\cM_\mu \,
X^{\cN\,i} \,,
\label{covscalar}
\ee
for the scalar fields. Their covariant field strengths follow from the
commutator of two covariant derivatives, {\it e.g.},  
\be
  [{\cal D}_\mu, {\cal D}_\nu]\, \phi^i =
g \,\GTh_{\cM\cN}\, F_{\mu\nu}^\cM \, X^{\cN i} \;,
\label{DDF}
\ee
and take the form 
\ba
\label{field-strength}
\Theta_{\cM\cN}\,F_{\mu\nu}^\cM = \Theta_{\cM\cN}\Big( 
\dd_{\mu}A_{\nu}^\cM -\dd_{\nu}A_{\mu}^\cM - 
g\, \hat f_{\cP\cQ}{}^{\cM} \, A_{\mu}^\cP A_{\nu}^\cQ\Big) 
\;.
\ea
The extra minimal couplings \Ref{covscalar} render the Lagrangian
invariant under local transformations \Ref{iso-trans1},
\Ref{subgroup} provided we assume the following transformation
behavior of the vector fields
\be
\Theta_{\cM\cN} \,\delta A_\mu^\cM =  \Theta_{\cM\cN} \left(-\partial_\mu
\Lambda^\cM +g\, \hat f_{\cP\cQ} {}^{\cM}
\,A_\mu^\cP\,\Lambda^\cQ  \right)\,.
\ee
However, they violate supersymmetry and the central question is whether new
terms in the supersymmetry variations and in the Lagrangian can be found such
as to regain this symmetry. It is at this point that the need arises to
include a CS term for the vector fields,
\ba
\CL_{\rm CS} &=& \ft{1}{4}\,\I g\,\varepsilon^{\mu\nu\rho} \,
A_\mu^{\cM} \,\GTh_{\cM\cN} \Big( \dd_{\nu} A_{\rho}^\cN - 
\ft13 g\, \hat f_{\cP\cQ}{}^{\cN} \, A_{\nu}^\cP A_{\rho}^\cQ\Big)  \;,
\label{CS}
\ea
and assume the following supersymmetry transformations,
\ba
\Theta_{\cM\cN}\,
\Gd A^\cM_\mu &=& \Theta_{\cM\cN}\, \Big[
2\,\CV^{\cM\,IJ}\, \Bpsi{}_\mu^I\Ge^J
+\CV^\cM{}_{i} \, \Bchi^{i I}\Gg_\mu\Ge^I \Big]  \;.
\label{susyA}
\ea
In order for this to work and to preserve gauge invariance, it is necessary to
adopt a symmetric, gauge invariant, embedding tensor. 

Although the embedding tensors must be found case by case, let us briefly 
mention some general properties. For semisimple gaugings, the Lie algebra
$\mathfrak{g}_0$ always decomposes as a direct sum 
\be\label{sumg}
\mathfrak{g}_0 = \bigoplus_i \mathfrak{g}_{i} \subset \mathfrak{g}
\;,
\ee
of simple Lie algebras $\mathfrak{g}_{i}$. In this case, the embedding tensor
can be written as a sum of projection operators  
\be\label{projector}
\Theta_{\cM\cN} = \sum_i \varepsilon_i 
\,\eta_{\cM\cP}  {(\Pi_i)^\cP}_\cN \;,
\ee
where $\Pi_i$ projects onto the $i$-th simple factor
$\mathfrak{g}_{0i}$, $\eta_{\cM\cP}$ is the Cartan-Killing form,
and the constants $\varepsilon_i$ characterize the relative strengths
of the gauge couplings. There is only one overall gauge coupling 
constant $g$ for the maximal theory ($N=16$), but there may be several 
independent coupling constants for lower $N$. 

For non-semisimple gaugings, \Ref{sumg} is replaced by 
\be\label{sumg1} 
\mathfrak{g}_0 =  \bigoplus_i \mathfrak{g}_{i} \oplus \mathfrak{t}
\;,
\ee
where $\mathfrak{t}$ represents the solvable part of the gauge
group.
 For the non-semisimple gauge groups which typically appear in
theories obtained by dimensional reduction, the latter subalgebra
decomposes into 
\be 
\mathfrak{t} = \mathfrak{t}_0 \oplus \mathfrak{t}'
\;.
\ee
The abelian subalgebra $\mathfrak{t}_0$ here transforms in the adjoint
of the semisimple part of the gauge group and pairs up with the
semisimple subalgebra in the embedding tensor, which has non-vanishing
components only in $(\mathfrak{g}_{i}\otimes \mathfrak{t}_0)_{\rm
sym}$ and in $(\mathfrak{t}'\otimes \mathfrak{t}')_{\rm sym}$. There are 
also many examples of nilpotent and almost nilpotent gaugings, where the
semisimple part is absent or `small'. Many examples of
non-semisimple gaugings can be generated from semisimple ones by a
singular ``boost'' within the global symmetry group ${\rm G}$, as
explained in
\cite{FiNiSa03}.

\section{$T$-tensors, consistency constraints, and the Lagrangian} 
Before presenting the full Lagrangian of the gauged supergravity, we define
the so-called $T$-tensor (originally introduced in higher-dimensional 
supergravity~\cite{dWiNic82}) as  
\ba
\begin{array}{rclrcl}
T^{IJ,KL} &\equiv& \CV^{\cM\,IJ}\GTh_{\cM\cN}\CV^{\cN\,KL}
\;,\qquad
&
T^{IJi} &\equiv&  \CV^{\cM\,IJ}\GTh_{\cM\cN}\CV^{\cN\,i}
\;,
\\[1ex]
T^{ij} &\equiv& \CV^{\cM\,i}\GTh_{\cM\cN}\CV^{\cN\,j}
\;,
&
T_\Ga{}^i &\equiv&  \CV^{\cM}{}_{\Ga}\GTh_{\cM\cN}\CV^{\cN\,i}
\;,
\\[1ex]
T_{\Ga\Gb} &\equiv& \CV^{\cM}{}_{\Ga}\GTh_{\cM\cN}\CV^{\cN}{}_{\Gb}
\;,
&
T^{IJ}{}_\Ga &\equiv&  \CV^{\cM\,IJ}\GTh_{\cM\cN}\CV^{\cN}{}_{\Ga}
\;. 
\end{array}
\label{T-tensors}
\ea 
The $T$-tensor components that carry indices $\alpha,\beta$ do not 
appear directly in the Lagrangian and transformation rules and are only
defined for $N>2$. From~\Ref{subgrouptheta} and \eqn{cov-CV} it readily
follows that the $T$-tensor transforms covariantly under the gauged
isometries. The additional masslike terms and the scalar potential in the
Lagrangian and the corresponding terms in the supersymmetry variations of the
fermion fields, which we will specify shortly, are 
encoded in three tensors, $A_1, A_2$ and $A_3$, which are related to the
$T$-tensor. 
  
A central result of \cite{dWHeSa03} is that a gauge group 
${\rm G}_0\subset{\rm G}$ with a gauge invariant embedding tensor 
$\Theta_{\cM\cN}$ describing the minimal couplings according to
\Ref{covscalar}, is consistent with supersymmetry if and only if the
associated $T$-tensor \Ref{T-tensors} satisfies the constraint, 
\ba
T^{IJ,KL} &=& T^{[IJ,KL]} 
- \frac4{N\mis2}\,\delta\oversym{^{I[K}\,T^{L]M,MJ}\;} 
- \frac{2\,\delta^{I[K}\delta^{L]J}}{(N\mis1)(N\mis2)}\,T^{MN,MN} \;.
\label{relT}
\ea
For $N=1$ and $N=2$, this constraint degenerates to an identity.  
The consistency constraint \Ref{relT} has a simple group-theoretical
meaning in $\SO{N}$: denoting the irreducible parts of $T^{IJ,KL}$ 
under $\SO{N}$ by  
\ba
\Yboxdim6pt
\left( \; {\young(\hfil,\hfil)} \;\times \, 
{\young(\hfil,\hfil)}\; \right)_{\rm sym}
~\;=\;~ 1 \;\;+\;\;
{\young(\hfil\hfil)} \;\;+\;\;
{\young(\hfil\hfil,\hfil\hfil) }
\;\;+\;\; 
{\yng(1,1,1,1) }\;\; ,
\label{IJKL}
\ea
with each box representing a vector representation of 
$\SO{N}$, 
equation \Ref{relT} eliminates the ``Weyl-tensor'' type representation
\ba
\Yboxdim3pt
\bbP\!_{\atop{}{\yng(2,2) }}\; 
T^{IJ,KL} &=& 0\;.
\label{con88}
\ea
The condition for a consistent gauging is now fully captured by constraints 
\eqn{subgrouptheta} and \eqn{relT} applied to a symmetric embedding tensor
$\Theta_{\cM\cN}$. Specific cases will be discussed in later
sections. 

Let us now present the full Lagrangian and transformation rules. The
Lagrangian is given by 
\ba
\CL &=& -\ft12\,\I\,\varepsilon^{\mu\nu\rho}\left(
e_\mu{}^a\,R_{\nu\rho a} +
\Bpsi{}^I_\mu {\cal D}_\nu\psi^I_\rho  \right)
-\ft12 e\,g_{ij} \left( g^{\mu\nu}\, {\cal D}_\mu \phi^i\,
{\cal D}_\nu \phi^j + N^{-1} \Bchi{}^{iI}\, \FMslash{{\cal D}}
\chi^{jI} \right)\non[1ex]
&&{}
+\ft{1}{4}\,\I g\,\varepsilon^{\mu\nu\rho} \,
A_\mu^{\cM} \,\GTh_{\cM\cN} \Big( \dd_{\nu} A_{\rho}^\cN - 
\ft13 g\, \hat f_{\cP\cQ}{}^{\cN} \, A_{\nu}^\cP A_{\rho}^\cQ\Big)
\non[1ex]
&&{}+ \ft14 e\, g_{ij}\,  \Bchi{}^{iI}\Gg^\mu\Gg^\nu
\psi_\mu^I\,({\cal D}_\nu\phi^j+\widehat {\cal D}_\nu \phi^j)  
-\ft1{24} e \,N^{-2} R_{ijkl} \; \Bchi{}^{iI} \gamma_a
\chi^{jI} \, \Bchi{}^{kJ} \gamma^a \chi^{lJ} \non[1ex] 
&&{} +  \ft1{48} e\,N^{-2} \left( 3\,( g_{ij}\, \Bchi{}^{iI}
\chi^{jI})^2 - 2(N\!-\!2)\, ( g_{ij}\, \Bchi{}^{iI}\gamma^a 
\chi^{jJ})^2  \right) \non[1ex]
&&{}+e g\Big(
\ft12  A_1^{IJ}\,\Bpsi{}^I_\mu\,\Gg^{\mu\nu}\,\psi^J_\nu\, +
 A^{IJ}_{2\,j}\,\Bpsi{}^I_\mu\,\Gg^\mu \chi^{j J} +
\ft12  A_{3\,}{}_{ij}^{IJ}\, \Bchi^{i I}\chi^{j J}\Big)
\non[1ex]
&&{} -2\, e g^2 \left( g^{ij}\, A_{2i}^{I\,J} A_{2j}^{I\,J}
-2N^{-1} A_1^{IJ}A_1^{IJ}  \right)
\;,
\label{Lgauge}
\end{eqnarray}
with covariant derivatives defined by 
\begin{eqnarray}
\CD_\mu\phi^i &=& \partial_\mu \phi^i +g\, \Theta_{\cM\cN} \,A^\cM_\mu \,
X^{\cN\,i} 
\;,\qquad
  \widehat \CD_\mu \phi^i ~=~ \CD_\mu \phi^i 
  -\ft12 \Bpsi{}^I_\mu \chi^{iI} 
\;,\non[1ex]
\CD_\mu \psi^I_\nu &=& \left( \partial_\mu +\ft12
\omega^a_\mu \gamma_a \right ) \psi^I_\nu  +
\dd_\mu\phi^i Q_i^{IJ} \,\psi^J_\nu
+ g\, \GTh_{\cM\cN} A_\mu^\cM\,\CV^{\cN\,IJ} \,\psi^J_\nu \;,
\non[1ex]
{\cal D}_\mu \chi^{iI} &=& 
\left( \partial_\mu +\ft12 \omega^a_\mu \gamma_a \right)\chi^{iI} 
+\partial_\mu \phi^j \left( \Gamma^i_{jk}\, \chi^{kI} +
Q_j^{IJ} \chi^{iJ} \right)\non[.5ex]
&&{}+ g\, \Theta_{\cM\cN} A_\mu^{\cM} 
\left( \delta^i_j\, \CV^{\cN\,IJ} -\delta^{IJ}g^{ik}\,
D_k\CV^\cN{}_j 
\right) \chi^{jJ} \;. 
\end{eqnarray}
The supersymmetry transformations read 
\begin{eqnarray}
\Gd e_\mu{}^a &=& \ft12\,\Be{}^I\Gg^a\,\psi^I_\mu 
\;,\non[.5ex]
\Gd A^\cM_\mu &=& 
2\,\CV^{\cM\,IJ}\, \Bpsi{}_\mu^I\Ge^J
+\CV^\cM{}_{i} \, \Bchi^{i I}\Gg_\mu\Ge^I \;,\non[.5ex]
\Gd \psi^I_\mu &=& {\cal D}_\mu \epsilon^I   - \ft18 g_{ij} \,\Bchi{}^{iI}
\gamma^\nu \chi^{jJ}\, \gamma_{\mu\nu} \,\epsilon^J - 
\Gd \phi^i \,Q_i^{IJ} \psi^J_\mu + g\,A_1^{IJ} \Gg_\mu\,\Ge^J \;,\non[.5ex]
\Gd \phi^i &=& \ft12\,\Be^I\,\chi^{iI} 
\non[.5ex]
\Gd \chi^{iI} &=& \ft12 \left(\delta^{IJ}{\bf 1} \mis f^{IJ}
\right)^i{}_{\!j}\; 
\FMslash{\widehat {\cal D}} \phi^j \, \epsilon^J -
\Gd \phi^j \left(
\Gamma^i_{jk}\,\chi^{kI} 
+ Q_j^{IJ}\chi^{iJ} \right)  -  g N \, A_{2}^{JiI} \,\Ge^J 
\;,
\label{susytrafG}
\end{eqnarray}
with
\ba
\CD_\mu \epsilon^I &=& \left( \partial_\mu +\ft12
\omega^a_\mu \gamma_a \right ) \epsilon^I   +
\dd_\mu\phi^i Q_i^{IJ} \,\epsilon^J 
+ g\, \GTh_{\cM\cN} A_\mu^\cM\,\CV^{\cN\,IJ} \,\epsilon^J \;.
\ea
The gauge transformations take the form
\ba
\label{gauge-trafo}
\Gd\phi^i      &=& g \,\Theta_{\cM\cN} \,\Lambda^\cM  X^{\cN i} \;,
\non[.5ex]
\Gd \psi^I_\mu &=& g\, \Theta_{\cM\cN} \,\Lambda^\cM  
{\cal V}^{\cN IJ} \,\psi_\mu^J - 
\Gd\phi^i\, Q_i^{IJ} \psi^J_\mu \;,\non[.5ex]
\Gd \chi^{iI}  &=& g\, \Theta_{\cM\cN} \,\Lambda^\cM  
 (\chi^{jI}\, D_j \CV^{\cN i}+ {\cal V}^{\cN IJ}\, \chi^{iJ}) 
 - \Gd \phi^j \left(\Gamma^i_{jk}\,\chi^{kI} 
+ Q_j^{IJ}\chi^{iJ} \right) \;,
\non[.5ex]
\Theta_{\cM\cN} \,\delta A_\mu^\cM &=&  \Theta_{\cM\cN} (- \partial_\mu 
\Lambda^\cM +g\, \hat f_{\cP\cQ} {}^{\cM} 
\,A_\mu^\cP\,\Lambda^\cQ )\,.
\ea

For $N>2$, the tensor $A_1$ is given by 
\ba
A_1^{IJ} &=& -\frac4{N\mis2}\,T^{IM,JM} +
\frac{2}{(N\mis1)(N\mis2)}\,\delta^{IJ}\,T^{MN,MN} \;.
\label{id0}
\ea
In the cases $N=1, 2$, this tensor is only partially determined as we 
shall describe in the next section. For all values of $N$ 
the tensors $A_2$ and $A_3$ are given functions of $A_1$ and the 
$T$-tensor \Ref{T-tensors}, 
\ba
A_{2\,i}^{I\,J} &=& 
{1\over N} \Big\{ D_i A_1^{IJ} + 2\,T^{IJ}{}_i 
\Big\}  \;,
\non[1ex]
A_{3\,}{}^{IJ}_{ij}&=&
{1\over N^2} \Big\{ -2\,D_{(i}D_{j)}A_1^{IJ} + g_{ij}\,A_1^{IJ} +
A_1^{K[I} \,f_{ij}^{J]K} \non 
&& \hspace{9mm} 
+2\, T_{ij} \, 
\delta^{IJ}  
- 4\, D_{[i} T^{IJ}{}_{j]}-   2\, T_{k[i} \,f^{IJk}{}_{j]}  
\Big\}  \;.
\label{id1}
\ea
{}Using \Ref{relT} and its derivatives, one may verify that these tensors
satisfy the symmetries implied by their appearance in \Ref{Lgauge}:
\ba
A_1^{IJ}&=&A_1^{JI}\;,\quad
\bbP_I^K{}_i^j\, A^{J\,K}_{2\,j} ~=~  A^{JI}_{2\,i} \;,\quad
A_{3\,}{}_{ij}^{IJ}~=~A_{3\,}{}_{ji}^{JI}~=~
\bbP_I^K{}_i^j\,A_{3\,}{}^{KJ}_{kj}
 \;.
\ea

\section{Discussion of low $N$ theories}
In this section we discuss the gauged supergravities for low values of $N$,
following \cite{dWHeSa03}. The cases $N=1,2$ are special because the tensor
$A_1$ 
entering the potential and the gravitino mass term is not uniquely determined  
by the conditions derived in the foregoing section, and thus in general 
is not expressible in terms of the $T$-tensor alone. This leaves 
the freedom for additional deformations (and thus scalar field potentials)  
which are {\em not} induced by gauging. The additional freedom for 
$N=1$ and $N=2$ appears via real and complex holomorphic superpotentials, 
respectively. For $N\geq 3$, on the other hand, all deformations  
correspond to gaugings.  

\mathon
\subsection{$N=1$}
\mathoff

In this case, the target space is a Riemannian manifold of 
arbitrary dimension $d$. The tensor $A_1$ has just one component, 
which is a gauge invariant function $F(\phi)$ on the target space, 
\be
  \GTh_{\cM\cN}X^{\cN i}\dd_i F = 0 \;.
  \label{N1C}
\ee
Reading off the values for $A_2$ and $A_3$ from (\ref{id1}), we obtain 
\ba
    A_1 &=& F\;, \qquad
    A_{2\,i} ~=~ \dd_i F \;, \qquad
    A_{3\,ij} ~=~ g_{ij} F -2\, D_i\dd_j F +
    2\, T_{ij} \;,
  \label{N1A123}
\ea
with $T_{ij}=X^{\cM}_i\, \GTh_{\cM\cN} X^{\cN}_j\,$. 
As a consequence, any subgroup of isometries can be 
gauged (for example, by choosing a constant function $F$). The 
gravitino $\psi_\mu$ is never charged under the  
gauge group, and the gauging is restricted to the matter sector. The case 
$\GTh_{\cM\cN}=0$ and $F\neq0$ corresponds to deformations
of the original theory that are not induced by gaugings. The scalar potential
$V$ is given by
\be
  V =  2g^2 \left( g^{ij}\,\dd_iF\dd_jF - 2 \,F^2 \right) \;,
\label{N1potential}
\ee
so that the function $F$ serves as the {\em real
superpotential}. Stationary points of $F$ define (anti-de Sitter)  
supersymmetric ground states. 
An off-shell version of these results 
for abelian gauge groups was recently given in~\cite{Gn}.

\mathon
\subsection{$N=2$}
\mathoff

The target space in this case is a K\"ahler manifold and may be conveniently parametrized
by $d/2$ complex coordinates and their conjugates,  
$(\phi^i, \Bphi{}^\Bi)$. Its metric and the $\SO2$ connection are given in
terms of the K\"ahler $K(\phi,\bar \phi)$ potential as  
$g_{i\bar\jmath} = \partial_i \partial_{\bar\jmath} K$, 
$Q_i \equiv Q^{12}_i = -\ft14{\I} \partial_i K$.
Any subgroup of the invariance group can be gauged. 
Partial results for abelian $N=2$ gaugings have been obtained in 
\cite{IzqTow95,DKSS99,AboSam01,BeHaSa02}. 

According to (\ref{isomf}) only holomorphic isometries of the target 
space can be extended to symmetries of the Lagrangian. Such isometries are
parameterized by holomorphic Killing vector fields 
$(X^i,X^{\bar\imath})$,
\ba 
\partial_\Bi X^j &=& 0\;, 
\qquad   
D_i X_{\bar\jmath} + D_{\bar\jmath} X_i = 0 \,. 
\nn 
\ea 
The second condition implies that the K\"ahler potential remains invariant
under the isometry up to a K\"ahler transformation. We write this special
K\"ahler transformation in terms of a {\it holomorphic} 
function~${\mathcal{S}}(\phi)$, {\it i.e.}, 
\begin{equation} 
  \delta K(\phi,\bar\phi) = - X^i\,\partial_i K - 
  X^{\bar\imath}\,\partial_{\bar\imath}  K  =  
4 \I \,\bigl( {\mathcal{S}} -\bar{\mathcal{S}} \bigr)  \,. 
  \label{special-kahler-trans} 
\end{equation} 
Equation (\ref{DXS}) may then be solved as 
\ben 
 \mathcal{V} ~\equiv~ \mathcal{V}^{12}= 
-\ft14{\I} (X^i \partial_iK - X^{\bar\imath} 
  \partial_{\bar\imath}K)  + {\mathcal{S}}^{12}  
= 
  - \ft12{\I} X^i \partial_i K + 2\, \mathcal{S} \,. 
\een 
For every generator $X^\cM$ of the invariance group we thus identify 
a holomorphic function~${\mathcal{S}}^\cM$, determined by 
\Ref{special-kahler-trans} up to a 
real constant. The particular transformation (which we   
denote with the extra label $\CM=0$),  
\be 
X^{0\, i}=0 \;,\qquad \CS^0 = \ft12  \;, \qquad \CV^0=1\;,  
\label{N2SO2} 
\ee 
constitutes a central extension of the isometry group and generates 
the $\SO2$ R-symmetry group that acts exclusively on the fermions. These 
symmetries play a role in the presence of FI terms. We refer to
\cite{dWHeSa03} for further details.
 
For the $T$-tensor, we introduce the notation 
\ba 
  T&\equiv& T^{IJ,IJ}= 2\, T^{12,12} \;,\qquad 
    T_i~\equiv~ T^{12}{}_{\!i} =\ft12 \I \partial_i T  
    \;. 
\ea 
The tensor $A_1^{IJ}$ is determined by \Ref{id0} 
\be 
A_1^{11}=-T-e^{K/2}\,\Re W \;,\quad\; 
A_1^{22}=-T+e^{K/2}\,\Re W \;,\quad\; 
A_1^{12}=A_1^{21}=e^{K/2}\,\Im W \;, 
\ee 
with an {\em holomorphic superpotential} $W(\phi)$,  which, because of gauge
  covariance, must satisfy 
\ba 
    \Theta_{{\cM}{\cN}} \bigl( X^{{\cN}i} D_i W  
    -2 \I\, \mathcal{V}^\cN W \bigr)~=~\Theta_{{\cM}{\cN}} 
   \bigl(X^{{\cN}i} \partial_i W   
    - 4 \I\, \mathcal{S}^\cN W \bigr)
    ~=~ 0 \,, 
\label{DW} 
\ea 
with the K{\"a}hler covariant derivative $D_i W \equiv \dd_i W + \dd_i 
K\,W$. The tensors $A_2$, $A_3$ follow from \Ref{id1}; for $A_2$ we find,  
\ba 
    A_{2\,i}^{1\,1} = -\I A_{2\,i}^{1\,2} =  -\ft12 (\partial_i T    
    + e^{K/2} D_i W) \,,  \quad
     A_{2\,i}^{21} = \I A_{2\,i}^{2\,2}= 
 \ft12 \I ( \partial_i T - e^{K/2} D_i W ) 
\;. 
\ea 
The scalar potential of the gauged theory is
given by
\be
  V =  g^2 \Big(
  4\,g^{i\Bi}\,\dd_iT\,\dd_\Bi T -
  4\,T^2  
 +   g^{i\Bi}\,e^K D_i W D_\Bi \overline{W}-
  4\,e^K \, |W|^2 \Big) 
  \;.
 \label{potentialN2}
\ee
Note that in three dimensions, the scalar potential contains terms quartic in the
moment map $\mathcal{V}$, since the $T$-tensor is quadratic in
$\mathcal{V}$. This is in contrast with {\it e.g.}\ four dimensions, where
the corresponding part of the scalar potential is quadratic in $\mathcal{V}$.

Analogous to the $N=1$ case, there are two kinds of supersymmetric 
deformations of the original theory. On the one hand, there are the
gaugings, which are completely characterized by an embedding tensor
$\Theta_{\cM\cN}$. The above analysis shows that there is no
restriction on the $T$-tensor, and therefore any subgroup of the
invariance group of the theory is an admissible gauge group, as long
as its embedding tensor satisfies~\Ref{subgrouptheta}. On the other
hand there are the deformations described by the holomorphic
superpotential $W$, which are not induced by a gauging. In case both
deformations are simultaneously present, their compatibility
requires~\Ref{DW}. Pure $N=2$ supergravity
(without gauging) can have a cosmological constant corresponding to a
constant $W$ and vanishing $T$. This implies that the gravitino mass
matrix is traceless. An alternative way to generate a cosmological
term in pure supergravity makes use of gauging the R-symmetry
group. In that case, $T$ equals a nonzero constant (equal to $\Theta_{00}$) 
and $W=0$; the
gravitino mass matrix is then proportional to the identity.  
The latter version has been considered in~\cite{IzqTow95}. 

\subsection{Some comments on $N=3$ and $N=4$ theories} 
For $N=3$, the target space is a quaternion-K\"ahler manifold. In this 
case, the consistency condition \Ref{relT} reduces to an identity 
such that any subgroup of isometries can be consistently 
gauged. 
For $N=4$ on the other hand, the target space is locally  
a product of two quaternion-K\"ahler manifolds of dimension~$d_\pm$.  
The almost-complex structures $f^{IJ}$ decompose into two sets of  
three almost-complex structures $f^\pm$,
\ba 
f^{+P} &\equiv& \ft12\,(J+ 1)\,f^P ~=~
\ft12 f^{P} - \ft14 \Ge^{PQR}\,f^{QR} \;,\qquad ( P=1,2,3)\;,\non 
f^{-\bar P} &\equiv& \ft12\,(J- 1)\,f^P ~=~
-\ft12 f^{P} - \ft14 \Ge^{PQR}\,f^{QR} \;,\qquad ( \bar{P}=1,2,3)\;,
\label{PIJ}
\ea
corresponding to the decomposition of the ${\rm SO}(4)$
R-symmetry group, 
\ba
\SO4 =  \SO3^+ \times \SO3^- \;.
\label{SO33}
\ea
In this basis, the consistency condition \Ref{relT} that encodes
supersymmetry of the theory takes the form
\ba
T^{P Q} &=& \ft13\,\delta^{P Q}\,T^{R R}
\;,\qquad\mbox{where}\quad
T^{P Q} ~=~
\CV^{\cM\,P}\GTh_{\cM\cN}\CV^{\cN\,Q}
\;,
\label{conN4}
\ea 
and correspondingly for $T^{\bar P\bar Q}$. The off-diagonal
components, $T^{P\bar Q}$ are constrained by the quadratic 
constraint~\Ref{subgrouptheta}, see reference~\cite{dWHeSa03} for details.
Unlike the cases $N<4$, it is thus no longer possible to gauge any subgroup 
of the isometry group. We refer to \cite{dWHeSa03} for further details.
Henceforth, we will call a subgroup of ${\rm G}$ `admissible' if its embedding
tensor obeys \Ref{embedding-cond} and \Ref{relT}, so that supersymmetry is  
preserved. 

\section{Symmetric target spaces with $N>4$}
Beyond $N=4$, the only admissible target spaces are the symmetric spaces 
listed in table~\ref{GG1}. Hence they are coset spaces ${\rm G}/{\rm H}$,
where the isotropy group is equal to the (maximal) holonomy group 
${\rm SO}(N)\times {\rm H}^\prime$. The scalar fields may be described by
means of a ${\rm G}$-valued matrix $\Mat(\phi^i)$, on which the rigid action
of ${\rm G}$ is realized by left multiplication, while ${\rm SO}(N)\times
{\rm H}^\prime$ acts as a local symmetry by multiplication from the right. The
generators of the group G constitute a Lie algebra $\mathfrak{g}$, which
thus decomposes into $\{ t^{\cM}\} = \{X^{IJ}, X^\alpha, Y^{A} \}$. The
$X^{IJ}$ generate $\SO{N}$, the $X^\alpha$
generate the compact group ${\rm H}^\prime$, while the remaining (noncompact)
generators $Y^A$ transform in a spinor representation of $\SO{N}$. 
The connection with the general quantities introduced above is 
given via 
\ba 
\Mat^{-1} \dd_i \Mat &=& \ft12\,Q_i^{IJ}\, X^{IJ} + 
Q^\alpha_i\, X^{\alpha} + e_i{}^A\, Y^{A} \;, 
\non 
X^{\cM\,i}\,\partial_i \Mat &=& t^\cM\, \Mat -  
\ft12 {\cal S}^{\cM IJ} \,\Mat\, X^{IJ}  
+ {\cal S}^{\cM \alpha}\, \Mat\, X^{\alpha} 
\;, 
\non 
\Mat^{-1} t^\cM \Mat &=&   
\ft12\,\CV^{\cM\,IJ}\, X^{IJ} + 
\CV^\cM{}_{\alpha}\, X^{\alpha} + \CV^\cM{}_{A}\, Y^{A} \;, 
\non  
g_{ij} &=& e_i{}^A\,e_j{}^B\,\delta_{AB} \;,\quad 
f^{IJ}_{ij} ~=~ - \Gamma^{IJ}_{AB}\,e_i^A\,e_j^B \;,\quad 
\CV^\cM{}_{i}~=~ e_i{}^A\, \CV^\cM{}_A 
\;.  
\label{hom} 
\ea  

Equations \eqn{ABCH} then correspond to the fact that the map 
\ba 
t^\cM &\rightarrow& \Mat^{-1} t^\cM \Mat \;, 
\label{iso}
\ea 
is an isomorphism of the algebra ${\mathfrak g}$; the actual equations follow
straightforwardly from the commutator $[\Mat^{-1} t^\cM \Mat,\, \Mat^{-1} t^\cN
\Mat]$, upon using the explicit commutation relations of the generators
$X^{IJ}$, $X^\alpha$ and $Y^{A}$. 
Linear first-order differential equations such as \Ref{DV} can be derived for
any coset space (see, e.g. \cite{dWit02}) and the actual results follow after 
substituting the appropriate expressions for the coset-space curvatures. Here
we should add that the above analysis can be straightforwardly extended to
$N=4$ with symmetric target spaces, as all these spaces are known and exhibit
the same characteristics as outlined above. 

The symmetric space structure in particular implies, that the $T$-tensor 
\Ref{T-tensors}
coincides with the image of the embedding tensor $\GTh_{\cM\cN}$ under
\Ref{iso}, 
\ba
T_{\cA\cB} &=& \CV^\cM{}_{\cA}\,\GTh_{\cM\cN}\,\CV^\cN{}_{\cB}
\;,
\label{Thom}
\ea
This allows us to lift the consistency condition \Ref{con88}, according to
which the $\SO{N}$ representation~$\Yboxdim4pt{\yng(2,2) }$  in the $T$-tensor
vanishes, to a field-independent condition on the embedding tensor, 
\be 
\label{Theta1} {\mathbb{P}_{\cM\cN}}^{\cP\cQ} \, 
\Theta_{\cP\cQ} = 0 \;.
\ee 
Here $\mathbb{P}$ projects onto the unique irreducible representation in 
$(\mathfrak{g}\otimes\mathfrak{g})_{\rm sym}$ that contains 
the~$\Yboxdim4pt{\yng(2,2) }$ representation of the $T$-tensor, via
\Ref{Thom}. 
It is a non-trivial fact that the $T$-tensor, which is assigned to R-symmetry 
representations, and appears in the fermionic masslike terms and the scalar
potential, can be 
assembled into representations  of the global symmetry group ${\rm G}$, as was
first noticed in the context of maximal gauged supergravity in four dimensions 
\cite{deWNic84}. 
For the symmetric target spaces, admissible subgroups of ${\rm G}$
are characterized by an embedding tensor that obeys \Ref{embedding-cond} and
\Ref{Theta1}. 

Note that the consistency conditions \Ref{embedding-cond} and \Ref{Theta1}
remain covariant under the {\em complexified} global  
symmetry group ${\rm G}_{\C}$. Indeed, non-semisimple gaugings in four 
dimensions were originally found in~\cite{Hull84a} 
by analytic continuation of $\SO8$ in the complexified global  
symmetry group ${\rm E}_7(\C)$. In three dimensions, a similar 
construction  should exist relating the different non-compact real 
forms of the  gauge groups listed in table~\ref{background} below, and 
explaining why ratios of coupling  constants between the factor groups 
remain the same.

\section{Admissible gauge groups for $N=16$} 
To illustrate the variety of possible gaugings, we now turn to the
maximally extended $N=16$ supergravity.\footnote{A different version of gauged N=16 supergravity, 
which modifies the
  ungauged theory only by topological terms, and does not lead to a
  scalar potential or Yukawa type couplings, was recently proposed in~\cite{Nn}.}  
  In this case the embedding 
tensor transforms as an element of the symmetric tensor product
of two adjoint (and in this case also fundamental) representations 
of ${\rm E}_{8(8)}$
\be
\big({\bf 248} \otimes {\bf 248}\big)_{\rm sym} 
=  {\bf 1} \oplus{\bf 3875} \oplus {\bf 27000}
\;,
\ee
As shown in \cite{NicSam00} \Ref{Theta1} becomes
\be\label{27000}
{\big(\mathbb{P}_{\bf 27000}\big)_{\cM\cN}}^{\cP\cQ} \,
\Theta_{\cP\cQ} = 0
\;.
\ee
so that the embedding tensor decomposes into a singlet and the ${\bf 3875}$
representations of ${\rm E}_{8(8)}$.
Following \cite{MarSch83}, we split the generators of 
$\mathfrak{g}=\mathfrak{e}_{8(8)}$ into 120 compact ones 
$X^{IJ}= - X^{JI}$ with $\SO{16}$ vector indices $I, J =1, \dots, 16$, 
and 128 noncompact ones $\{Y^A\}$ with $\SO{16}$ spinor indices 
$A=1, \dots 128$. Then the condition \Ref{27000} implies that only 
special $\SO{16}$ representations can appear in $\GTh$; we have
\be
\Theta = \GTh_{IJ|KL}\, X^{IJ} \otimes X^{KL} + 
      \GTh_{IJ|A} \, \big(X^{IJ} \otimes  Y^A + Y^A \otimes X^{IJ}\big) 
      + \GTh_{A|B}\, Y^A \otimes Y^B
\;,
\ee 
with \cite{NicSam00,FiNiSa02}
\ba\la{Theta}
\GTh_{IJ|KL} &=& -2 \Gth\, \Gd^{IJ}_{KL} +
  2 \Gd\undersym{_{I[K} \, \Xi_{L]J} \, } + \Xi_{IJKL}  \;, \non[1ex]
\GTh_{IJ|A}&=& - \ft17 \,\GG^{[I}_{A\dA}\, \Xi^{J]\dA}   \;, \non[1ex]
\GTh_{A|B} &=& \Gth \,\Gd_{AB} + \ft1{96} \,\Xi_{IJKL} \,\GG^{IJKL}_{AB}
\;,
\ea
and the $\SO{16}$ $\GG$ matrices $\GG^I_{A\dA}$, where the indices
$\dA=1, \dots, 128$ label the conjugate spinor representation. 
The tensors $\Xi_{IJ}$, $\Xi_{IJKL}$ and $\Xi^{I\dA}$ transform as
the $\bf{135}$, $\bf{1820}$ and $\bf\overline{1920}$ representations 
of $\SO{16}$, respectively; hence $\Xi_{II} = 0=\GG^I_{A\dA}\,
\Xi^{I\dA}$, and $\Xi_{IJKL}$ is completely antisymmetric in its 
four indices. The singlet contribution in \Ref{Theta} is absent for 
non-semisimple and complex gauge groups.

Although the solutions to \Ref{27000} have not been exhaustively
classified, it is known that all the irreducible components occurring 
in \Ref{Theta} can and do appear, depending on the type of gauge group.
The simplest examples are the semisimple gaugings with maximal 
supersymmetry constructed in \cite{NicSam00}, for which we have 
quite generally
\be
\theta \, , \Xi_{IJ}\, ,\, \Xi_{IJKL} \neq 0 \quad \mbox{and} \quad
\Xi^{I\dA}= 0 \qquad \mbox{(for semisimple $\mathfrak{g}_0$)}
\;.
\ee
In this case, the sum~\Ref{projector} contains at most two terms, 
i.e.\ the gauge groups are typically products of two simple groups 
${\rm G}_1\times {\rm G}_2$ with a fixed ratio of coupling constants 
$g_1/g_2$, such that there is only one free parameter in the theory. 
Schematically, we have the admissible gauge groups
\be\label{G0} 
G_0 = E_8 \, , \, E_7 \times A_1 \, , \, E_6 \times A_2 \, , \, 
F_4 \times G_2 \, , \, D_4 \times D_4 \;.
\ee 
which appear in all those real forms that are consistent with $E_{8(8)}$.
Remarkably, the ratio $g_1/g_2$ does not depend on the chosen real 
form. Furthermore, as shown in \cite{NicSam00,FiNiSa02}, all these
theories possess maximally supersymmtric (AdS or Minkowski) ground
states. The corresponding theories with their corresponding gauge groups,
which are particular noncompact versions of the groups \Ref{G0}, are 
listed in table~\ref{background}. In the last column, the table 
lists the symmetry groups of the ground states, which are 
superextensions of the three-dimensional AdS group 
${\rm SL}(2,{\mathbb R}) \times {\rm SL}(2,{\mathbb R})$. Besides the
fully supersymmetric vacua, there are also many known stationary points 
with partially broken supersymmetry \cite{Fisc02,FiNiSa02,Fisc03}.
However, because no general and complete results on the extremal 
structure of the associated potentials are available to 
date\footnote{Even for $D\!=\! 4$, the complexity of the potentials
has prevented the identification of new stationary points beyond
those already found in \cite{Warn83,HulWar85}, although the potentials 
are now known on a larger manifold of scalar field configurations 
thanks to the high performance symbolic algebra program developed 
in \cite{Fisc03}.}, many further extremal points could exist besides 
the known ones.

\begin{table}[t] 
\begin{center}
\vspace{2ex} 
\begin{tabular}{||c|c|c|c||}  
\hline 
&&&\\[-2.1ex]
gauge group ${\rm G}_0$
&
ratio $g_1/g_2$ 
&
$(n_{\rm L},n_{\rm R})\!$
&
ground state symmetry group\\[.5ex]
\hline
\hline
&&&\\[-2.1ex]
${\rm SO}(8)\cro {\rm SO}(8)$
& $g_1/g_2=-1$
&
$(8,8)$ 
& ${\rm OSp}(8|2,{\mathbb R})\cro {\rm OSp}(8|2,{\mathbb R})$\\[.5ex]
\hline
&&&\\[-2.1ex]
${\rm SO}(7,1)\cro {\rm SO}(7,1)$
& $g_1/g_2=-1$&
$(8,8)$ 
& 
${\rm F}(4)\cro {\rm F}(4)$\\[.5ex]
\hline
&&&\\[-2.1ex]
${\rm SO}(6,2)\cro {\rm SO}(6,2)$
 & $g_1/g_2=-1$&
$(8,8)$
&
${\rm SU}(4|1,1)\cro {\rm SU}(4|1,1)$\\[.5ex]
\hline
&&&\\[-2.1ex]
${\rm SO}(5,3)\cro {\rm SO}(5,3)$
& $g_1/g_2=-1$
&
$(8,8)$ &
${\rm OSp}(4^*|4)\cro {\rm OSp}(4^*|4)$\\[.5ex]
\hline
&&&\\[-2.1ex]
${\rm SO}(4,4)\cro {\rm SO}(4,4)$
& $g_1/g_2=-1$
&
$(8,8)$ &
Minkowski vacuum\\[.5ex]
\hline
&&&\\[-2.1ex]
${\rm G}_{2(2)}\!\times\!{\rm F}_{4(4)}$ 
& $\cc{{\rm G}_2}/\cc{{\rm F}_4}=-3/2$&
$(4,12)$ &
${\rm D}^1(2,1;-\ft23) \cro {\rm OSp}(4^*|6)$ \\[.5ex]
\hline
&&&\\[-2.1ex]
${\rm G}_{2}\!\times\!{\rm F}_{4(-20)}$ 
 & $\cc{{\rm G}_2}/\cc{{\rm F}_4}=-3/2$ &
$(7,9)$&
${\rm G}(3)\cro {\rm OSp}(9|2,{\mathbb R})$ \\[.5ex]
\hline
&&&\\[-2.1ex]
${\rm E}_{6(6)}\!\times\!{\rm SL}(3)$
&
$\cc{{\rm A}_2}/\cc{{\rm E}_6}= -2$&
$(16,0)$&
${\rm OSp}(4^*|8)\cro {\rm SU}(1,1)$ 
\\[.5ex] 
\hline
&&&\\[-2.1ex]
${\rm E}_{6(2)}\!\times\!{\rm SU}(2,1)$
&
$\cc{{\rm A}_2}/\cc{{\rm E}_6}= -2$&
$(12,4)$
&
${\rm SU}(6|1,1) \cro {\rm D}^1(2,1;-\ft12) $
\\[.5ex] \hline
&&&\\[-2.1ex]
${\rm E}_{6(-14)}\!\times\!{\rm SU}(3)$
&
$\cc{{\rm A}_2}/\cc{{\rm E}_6}= -2$&
$(10,6)$&
${\rm OSp}(10|2,{\mathbb R})\cro {\rm SU}(3|1,1)$
\\[.5ex] 
\hline
&&&\\[-2.1ex]
${\rm E}_{7(7)}\!\times\!{\rm SL}(2)$  
& $\cc{{\rm A}_1}/\cc{{\rm E}_7}= -3 $& 
$(16,0)$ 
 & 
${\rm SU}(8|1,1)\cro {\rm SU}(1,1)$  
\\[.5ex]
\hline
&&&\\[-2.1ex]
${\rm E}_{7(-5)}\!\times\!{\rm SU}(2) $ 
 & $\cc{{\rm A}_1}/\cc{{\rm E}_7}= -3 $&
$(12,4)$
&
${\rm OSp}(12|2,{\mathbb R}) \cro {\rm D}^1(2,1;-\ft13)$
\\[.5ex]
\hline
&&&\\[-2.1ex]
${\rm E}_{8(8)}$
&
$\cc{{\rm E}_8}$&
$(16,0)$
&
${\rm OSp}(16|2,{\mathbb R})\cro {\rm SU}(1,1)$ \\[.5ex]
\hline
\end{tabular}
\caption{\small The $N=16$ theories with semisimple gauge groups $G_0$. Except for
the last row, the gauge groups appear as direct products of two factors whose
coupling constant ratio $g_1/g_2$ is determined by \Ref{27000}. All these
theories admit  a maximally supersymmetric AdS (or Minkowski, for 
${\rm G}_0={\rm SO(4,4)}\times {\rm SO(4,4)}$) ground  
state, whose symmetry group factorizes according to ${\rm G}_{\rm L}\times{\rm
  G}_{\rm R}$, as specified in the last column; the supercharges split
accordingly into $n_{\rm   L}+n_{\rm R}=16$.} 
\label{background}       
\end{center}
\end{table}

A second class are the non-semisimple gaugings, whose existence can also
be inferred from the fact that in higher dimensions there are 
many maximal gaugings with non-semisimple groups
\cite{Hull84a,ACFG00,ADFL02,Hull02,dWSaTr02}. For the 
non-semisimple gaugings, in general all components of the embedding 
tensor in \Ref{Theta} are non-vanishing, in particular the `off-diagonal'
components (mixing compact and non-compact generators)
\be
\Xi^{I\dA}\neq 0 \qquad \mbox{(for non-semisimple $\mathfrak{g}_0$)}
\;.
\ee
For $N=16$, the most prominent examples are \cite{FiNiSa03}
\ba\label{G1}
{\rm G}_0 &=& \SO{{p,q}} \ltimes {\rm T}_{28} 
                \quad \mbox{ for $p+q=8$}\; ;  \nonumber\\
{\rm G}_0 &=& {\rm CSO}(p,q;r) \ltimes {\rm T}_{p,q,r} 
        \quad \mbox{for  $p+q + r=8$ and  $r>0$} 
\ea
Here, ${\rm T}_{28}$ is an abelian group of 28 translations transforming 
in the adjoint of ${\rm SO}(p,q)$. Similarly, ${\rm T}_{p,q,r}$ is a 
group of translations, but of smaller dimension
\be\label{CSO}
{\rm dim} \,  {\rm T}_{p,q,r} = {\rm dim} \, {\rm CSO}(p,q;r) = 28 -
\ft12 r(r- 1)
\;.
\ee
Note that the groups in \Ref{G1} involving ${\rm SO}(p,q)$ or 
${\rm CSO}(p,q;r)$ with $p\not=0,8$ admit {\em only one} embedding,
whereas there are {\it two} inequivalent $SO(8)\ltimes T_{28}$ gaugings,
corresponding to the compactifications IIA and IIB supergravity 
on $S^7$. Quite generally, reduction of a higher-dimensional gauged  
supergravity (with semisimple or non-semisimple gauge group) on a
torus will always lead to a non-semisimple gauge group in three 
dimensions. In view of the equivalence of CS and YM type gauge
theories explained in section~4, the gauged supergravities with
the gauge groups \Ref{G1} are consequently on-shell equivalent to
the ones obtained by reducing the $SO(p,q)$ and $CSO(p,q;r)$ theories 
of \cite{Hull84a} on $S^1$. Further examples of non-semisimple gaugings 
can be generated from semisimple ones by the boost method described 
in \cite{FiNiSa03}.

In contrast to the semisimple gaugings, the non-semisimple
ones do not admit maximally supersymmetric groundstates. The potentials
contain exponential factors and their minimum is usually reached at
infinity. This phenomenon is well-known from higher-dimensional gauged
supergravities. The non-existence of fully supersymmetric vacua is also  
related to the disappearance of the supersymmetric vacuum that 
is known to occur when one reduces maximal gauged supergravity from 
four or five to three dimensions on a torus.

The most curious solution of the consistency conditions is the
{\em complex} gauge group 
\ba\label{G3}
{\rm G}_0 &=& {\rm SO}(8,\C)
\;.
\ea
which can be realized {\em in two inequivalent ways}, corresponding to 
two possible and inequivalent embeddings of ${\rm SO}(8,\C)$ into
the (real) Lie group ${\rm E}_{8(8)}$ (there are similar 
complex gauge groups ${\rm SO}(n,\C)$ for $N=2n=12,10$
supergravities). This gauging provides an example of a purely 
off-diagonal embedding tensor for which 
\be
\Gth = \Xi_{IJ}=\Xi_{IJKL} = 0 \quad \mbox{and} \quad \Xi^{I\dA} \neq 0
\qquad 
\mbox{(for $\mathfrak{g}_0 = \mathfrak{so}(8,\C)$)} \;,
\ee
so $\Xi^{I\dA}$ is the {\em only} nonvanishing component in \Ref{Theta}. 
Because it does not require an imaginary unit, this embedding 
exhibits some rather strange properties. Like the semisimple 
gauge groups of table~\ref{background}, the ${\rm SO}(8,\C)$ gauged
supergravities cannot be derived from higher dimensions by any known
mechanism. Furthermore, they feature a de Sitter stationary point at 
the origin breaking all supersymmetries, and with tachyonic 
instabilities. (There are indications that these models possess no 
further extrema besides the one at the origin.\footnote{T.~Fischbacher,
private communication.}) We note that CS gauge theories with complex 
gauge groups are of considerable interest (\cite{Witt91}; see also 
\cite{Guko03} and references therein for some recent
developments). The embedding of such theories into supergravity
with non-trivial matter couplings may well provide interesting
new perspectives.

As we already explained in the introduction, the existence of the large
variety of gauged supergravities in three space-time dimensions, with 
potentials that have stationary points corresponding to AdS backgrounds, is
important in the context of the AdS/CFT correspondence. 
In the case at hand the correspondence implies a relation between 
an AdS solution of a certain three-dimensional gauged supergravity and a
two-dimensional (super)conformal theory living on the boundary of the AdS
space.  
The two-dimensional theories are characterized by an infinite-dimensional
superconformal algebra. These algebras have all been classified
\cite{classification}; they consist of a sum of 
two algebras, pertaining to the left- and right-moving sectors,
respectively, containing an $n_{\rm L}$- and an $n_{\rm R}$-superextended 
Virasoro algebra. On the supergravity side, the maximal
finite-dimensional subalgebra will correspond to the symmetry algebra 
of the ${\rm AdS}_3$ stationary point. To illustrate this,
one may consider the theories listed in table~\ref{background}, which admit
maximally supersymmetric ${\rm AdS}_3$ stationary points whose symmetry
algebra are listed in the last column. Indeed, each of these symmetry algebras
coincides with the maximal finite subalgebra of a corresponding superconformal
algebra of \cite{classification} 
with the appropriate numbers, $n_{\rm L}$ and $n_{\rm R}$, of supercharges.

The infinite-dimensional superconformal algebras appear in the asymptotic
symmetries of the supergravity fields~\cite{BroHen86,BBCHO98}. For the pure  
extended ($N>1$) supergravity theories, this phenomenon was analyzed
in~\cite{HeMaSc99}. For $n_{\rm L,R}>4$, this analysis confirmed the presence
of terms in the algebra that are quadratic in the generators, in accord with
the known form of the corresponding infinite-dimensional superconformal
algebras. It should be interesting to extend this analysis to the propagating
bulk fields described by the matter-coupled  gauged supergravities of this
paper.  

In the spirit of the AdS/CFT correspondence 
the supergravity Lagrangians~\Ref{Lgauge} obtained for the theories
listed in table~\ref{background} allow the construction of the $n$-point
correlation functions of a closed subset of chiral primary operators
of the associated superconformal theories.
To date, no concrete proposal for these $N=16$ boundary theories 
has been  put forward --- partly due to the lack of known brane 
configurations whose near horizon geometry would 
admit an isometry group related to any of the gauge groups in 
table~\ref{background}. In contrast, the most prominent example
of an AdS$_3$/CFT$_2$ correspondence, the D1-D5 system, relates
IIB string theory on ${\rm AdS}_3\times S^3\times M_4$ \cite{stringAdS3} 
to an $N=(4,4)$ superconformal field theory described 
by a non-linear sigma model whose target space is a deformation of 
the symmetric orbifold $(M_4)^n/S_n$ \cite{AGMOO99}. The corresponding
low-energy effective supergravity is the half-maximal theory constructed 
in~\cite{NicSam03b}.
 
Gauged supergravities with non-semisimple gauge groups on the other
hand make their appearance in the generalization of the AdS/CFT 
correspondence to so-called domain wall/QFT dualities, 
relating string theory on near-horizon Dp-brane geometries to
$d=p\pls1$ dimensional super-Yang-Mills theories with sixteen
supercharges~\cite{IMSY98,BoSkTo98}. In particular, the $N=16$ theory
with gauge group $\SO{{8}} \ltimes {\rm T}_{28}$
describing the warped ${\rm AdS}_3\times S^7$ near-horizon
D-string geometry~\cite{MorSam02}, is holographically dual
to IIA matrix string theory~\cite{DiVeVe97}.

\subsection*{Acknowledgments}
B.~de Wit and H.~Nicolai would like to thank the organizers for
a very enjoyable meeting. 
We thank T. Fischbacher and I. Herger for collaboration on the
results reported here. This work is partly supported by EU contracts
HPRN-CT-2000-00122 and HPRN-CT-2000-00131, and by the INTAS 
contract 99-1-590.

\bigskip
\bigskip

\providecommand{\href}[2]{#2}\begingroup\raggedright\endgroup

\end{document}